\documentclass[preprint]{elsarticle}
\pdfoutput=1
\usepackage{geometry,graphicx,amsmath,amssymb,amsfonts,mathdots,mathrsfs,caption,subcaption, setspace, wrapfig}
\usepackage[usenames,dvipsnames,svgnames,table]{xcolor}
\usepackage{graphicx}
\usepackage{braket}
\usepackage{comment}
\usepackage{hyperref}
\usepackage{slashed}
\usepackage{hyperref}
\biboptions{sort&compress}
\usepackage{epsfig}
\usepackage{longtable}
\usepackage{enumitem}

\journal{Physics Letters B}

\newcommand{\be}{\begin{equation}}
\newcommand{\ee}{\end{equation}}
\newcommand{\bea}{\begin{eqnarray}}
\newcommand{\eea}{\end{eqnarray}}
\newcommand{\bwt}{\begin{widetext}}
\newcommand{\ewt}{\end{widetext}}

\begin{document}

\begin{frontmatter}
\title{Vector-like Quarks with a Scalar Triplet}
\author[loc:uba]{Estefania Coluccio Leskow\fnref{em:tif}}
\author[loc:triumf]{Travis A. W. Martin\fnref{em:tam}}
\author[loc:triumf]{Alejandro de la Puente\fnref{em:adlp}}

\fntext[em:tif]{Email: ecoluccio@df.uba.ar}
\fntext[em:tam]{Email: tmartin@triumf.ca}
\fntext[em:adlp]{Email: adelapue@triumf.ca}

\address[loc:uba]{CONICET, IFIBA, Departamento de F\'{\i}sica, FCEyN, Universidad de Buenos Aires,\\ Ciudad Universitaria, Pab.~1, (1428), Ciudad de Buenos Aires, Argentina.}
\address[loc:triumf]{TRIUMF, 4004 Wesbrook Mall, Vancouver, BC V6T 2A3, Canada}

\begin{abstract}

We study a minimal extension to the Standard Model with an additional real scalar triplet, $\Sigma$, and a single vector-like quark, $T$. This class of models appear naturally in extensions of the Littlest Higgs model that incorporate dark matter without the need of $T$-parity. We assume the limit that the triplet does not develop a vacuum expectation value and that all dimension five operators coupling the triplet to Standard Model fields and the vector-like quarks are characterized by the scale $\Lambda$ at which we expect new physics to arise. We introduce new non-renormalizable interactions between the new scalar sector and fermion sector that allow mixing between the Standard Model third generation up-type quark and the vector-like quark in a way that leads to the cancellation of the leading quadratic divergences to the one-loop corrections from the top quark to the mass of the Higgs boson. Within this framework, new decay modes of the vector-like quark to the real scalar triplet and SM particles arise and bring forth an opportunity to probe this model with existing and future LHC data. We contrast constraints from direct colliders searches with low energy precision measurements and find that heavy vector-like top quarks with a mass as low as $650$ GeV are consistent with current experimental constraints in models where new physics arises at scales below $2$ TeV.

\end{abstract}
\begin{keyword}
Top Partner \sep Vector-Like Quark \sep Scalar Triplets \sep Naturalness \sep Electroweak Precision Constraints \sep Direct Collider Constraints
\end{keyword}
\end{frontmatter}

\section{Introduction}\label{sec:Introduction}

Energies beyond the electroweak scale are now being probed by the LHC, and searches for new particles and interactions are now underway. While the discovery of the Higgs boson was a primary goal of the LHC~\cite{Aad:2012tfa,Chatrchyan:2013lba}, many believe that the resolution to the electroweak hierarchy problem should also be discoverable with the LHC. The ultraviolet sensitivity of the Higgs mass provides a strong motivation for physics at the TeV scale. 

Within extensions of the Standard Model (SM), vector-like quarks are one attractive scenario to address the top quark contribution to the hierarchy problem. Within little Higgs models~\cite{ArkaniHamed:2002qy,Perelstein:2003wd}, new vector-like quarks are related to the SM chiral fields via symmetries and contribute to electroweak symmetry being broken collectively, where multiple operators act to protect the Higgs boson mass from developing a quadratic divergence. These operators also appear in light composite Higgs models~\cite{Contino:2006qr,Matsedonskyi:2012ym}, where the Higgs is a pseudo nambu-Goldstone boson with a potential generated by top quark loops with exotic coloured quarks appearing in the spectrum. Furthermore, extended symmetries that lead to this behaviour often also include new scalar fields, which need to be accounted for when applying precision and direct collider constraints to the models~\cite{Schmaltz:2010ac,Martin:2013fta}. Vector like quarks at the TeV scale are also strongly motivated in models where SM particles propagate in the bulk of an extra dimension~\cite{Gherghetta:2000qt,Grossman:1999ra,Huber:2000ie,Huber:2003tu,Agashe:2003zs,Agashe:2004bm,Agashe:2004cp,Agashe:2006at}. 

In our study, we focus on a scenario of a toy model with an additional real scalar triplet, $\Sigma$, and a single vector-like quark, $T$. We assume that the triplet does not develop a vacuum expectation value ($vev$), and that all dimension five operators are parametrized by the scale $\Lambda$ at which we expect new physics to arise. Within our model, a vector-like top quark is supplemented to cancel the quadratic divergence to the Higgs mass that arise from top quark loops. Furthermore, additional decay modes of the vector-like top quark to the real scalar triplet and SM particles arise and bring forth an opportunity to probe this model with existing and future LHC data. This scenario is realized naturally in extensions of Little Higgs models aimed at providing a dark matter relic and decoupling the existence of TeV scale fermions from electroweak precision constraints~\cite{Schmaltz:2010ac,Martin:2013fta}. There exist two global symmetries within this class of models, $G_{\Delta}/H_{\Delta}$ and $G_{\Theta}/H_{\Theta}$, with the same gauged subgroup and with different breaking scales, $F$ and $f$, respectively. A crucial property of this setup is that only one combination of pseudo Nambu-Goldstone fields from the $\Theta$ and $\Delta$ sectors become the longitudinal component of the heavy gauge bosons. The orthogonal combination are physical degrees of freedom that couple to fermions and give rise to interesting phenomenology, such as new decay modes of heavy vector-like quarks.

Many precision constraints on our scenario and those mentioned above depend strongly on the values of coupling parameters in the model (often parametrized in terms of mixing angles), and place an upper limit on the value of the energy scales involved. Direct collider searches present an excellent complimentary search method by placing a lower limit on the same parameters - bounding the data by complimentary regions. In particular, we focus on three constraining measurements to study the parameter space of our model - the oblique parameters ($S$, $T$, and $U$)~\cite{Peskin:1991sw}, $Z\rightarrow \bar{b}b$~\cite{Baak:2012kk}, and searches from the LHC for heavy, vector-like quarks~\cite{cmsHT,atlasHT}. 

This paper is organized as follows: in Section \ref{sec:ToyModel}, we discuss the scalar sector of the toy model; we start by presenting the extension of the sector with the addition of a real scalar triplet, while in Section \ref{sec:tripletvector} we discuss the phenomenology that emerges from the implementation of a vector like top quark. In Section \ref{sec:constraints} we study the constraints to our model, while in Section \ref{sec:Results} we present the results. In Section~\ref{sec:Conclusions} we provide some concluding remarks.

\section{Toy Model}\label{sec:ToyModel}

\subsection{Real Scalar Triplet}

The possibility of extending the SM with a real $SU(2)_{W}$ triplet scalar has been extensively studied~\cite{Gunion:1989ci,Blank:1997qa,Forshaw:2001xq,Forshaw:2003kh,Chen:2006pb,Chankowski:2006hs,SekharChivukula:2007gi,0811.3957,1303.4490,Brdar:2013iea} since such extensions generally lead to suppressed contributions to electroweak precision observables (EWPO). The scalar Lagrangian for a toy model including all possible gauge invariant combinations of a Higgs doublet, $H$, and an $SU(2)_{W}$ triplet, $\Sigma$, given by
\be
H = \left(\begin{array}{c}\phi^{+} \\ \phi^{0} \end{array}\right),  \quad \quad \quad \quad \quad \quad
\Sigma =\frac{1}{2}\left(\begin{array}{cc} \eta^{0} & \sqrt{2} \eta^+ \\ \sqrt{2} \eta^- & -\eta^{0} \end{array}\right), \label{eq:ScFields}
\ee
can be written as
\be
{\cal L}_{\mathrm{scalar}}=(D_{\mu}H)^{\dagger}(D^{\mu}H)+Tr(D_{\mu}\Sigma)^{\dagger}(D^{\mu}\Sigma)-V(H,\Sigma), \label{L}
\ee
where 
\begin{eqnarray}
V(H,\Sigma)&=&-\mu^2 H^{\dagger} H+\lambda_0(H^{\dagger} H)^2+\frac{1}{2}M^2_{\Sigma}Tr[\Sigma^2]+\frac{b_4}{4}Tr[\Sigma^2]^2\cr
&&+a_1H^{\dagger}\Sigma H+\frac{a_2}{2} H^{\dagger}H Tr[\Sigma^2], \label{V}
\end{eqnarray}
is the scalar potential~\cite{1303.4490,0811.3957}, and the covariant derivatives are the standard $SU(2)_{W}\times U(1)_{Y}$, as in the SM.

The scalar potential can be minimized along the directions of the neutral components of both $H$ and $\Sigma$, leading to two conditions:
\begin{eqnarray}
\frac{\partial V}{\partial Re( \phi^{0})}&=&\left(-\mu^{2}+\lambda_{0}v^{2}_{0}+\frac{a_{1}v_{3}}{2}+\frac{a_{2}v^{2}_{3}}{2}\right)v_{0}=0, \nonumber \\
\frac{\partial V}{\partial \eta^{0}}&=&M^{2}_{\Sigma}v_{3}+b_{4}v^{3}_{3}+\frac{a_{1}v^{2}_{0}}{4}+\frac{a_{2}v^{2}_{0}v_{3}}{2}=0,\label{eq:minima}
\end{eqnarray}
where $v_{0}\equiv \left<Re(\phi^{0})\right>$ and $v_{3}\equiv \left<\eta^{0}\right>$ are the vacuum expectation values, $vev$, of the neutral components of the SM complex doublet and the real triplet, respectively. This potential results in a mixing between the neutral and charged states, respectively, parametrized by mixing angles given by:
\begin{eqnarray}
\tan2\theta_{0}&=&\frac{4v_0 v_{3}(-a_{1}+2v_{3}a_{2})}{8\lambda_{0}v^{2}_3 v_{3}-8b_{4}v^{3}_{3}-a_{1}v^{2}_0}, \cr
\tan2\theta_{+}&=&\frac{4v_0 v_{3}}{4v^{2}_{3}-v^{2}_{0}}.
\end{eqnarray}
In the limit of $a_1 \rightarrow 0$, and for $M_\Sigma^2,b_4 > 0$, the minima of the scalar potential occurs at $v_3 \rightarrow 0$ and $v_0 \rightarrow v_{SM} \equiv v$, and there is no mixing between similarly charged components of the complex doublet and real triplet at tree-level. This represents an accidental $Z_2$ symmetry, as the potential remains invariant under the transformation $\Sigma \rightarrow -\Sigma$. 

In the most general scenario, the triplet $vev$ is non-zero, there is no $Z_2$ symmetry, and mixing does occur. This leads to contributions to the $\rho$ parameter proportional to $(v_3/v_0)^2$, and a constraint that $(2v_3/v_0)^2<0.001$, or $v_3 < 4$~GeV~\cite{0811.3957}. Taking $b_4$ and $a_2$ to be $\mathcal{O}(1)$, with $M_\Sigma$ to be $\mathcal{O}(100)$~GeV, this translates to a constraint on $a_1$ to be $\mathcal{O}(10)$~GeV. Since we are free to choose $a_1$ independent of other parameters. We choose the limit that $a_1 \rightarrow 0$ and neglect the triplet $vev$, as these issues are covered in great detail in~\cite{0811.3957}. This limit corresponds to a SM-like Higgs boson.

In the limit of no mixing, this toy model represents the addition of an inert scalar triplet to the SM, and the SM-like Higgs boson, $h^{0} \equiv Re(\phi^{0})$, acquires a mass as in the SM, given by
\begin{equation}
 M^{2}_{h^{0}}=2\lambda_{0}v^{2}.
\end{equation}
The real triplet masses are degenerate at tree level, given by
 \begin{equation}
 M^{2}_{\eta^{0}}=M^{2}_{\eta^{\pm}}=\frac{a_{2}v_0^{2}}{2}+M^{2}_{\Sigma} \equiv M_{\eta}^2.
 \end{equation}
This degeneracy will be broken by radiative corrections arising from the coupling between the triplet and the $SU(2)_W$ gauge bosons~\cite{Cirelli:2005uq,0811.3957}, 
resulting in a mass splitting of
\begin{equation}
\Delta M=\frac{\alpha M_{\eta}}{4\pi s^{2}_{W}}\left[f\left(\frac{M_{W}}{M_{\eta}}\right)-c^{2}_{W}f\left(\frac{M_{Z}}{M_{\eta}}\right)\right], 
\end{equation}
where the functions $f(M_W/M_\eta)$ and $f(M_Z/M_\eta)$ are given by
\begin{equation}
f(y)=-\frac{y}{4}\left[2y^{3}\log y +(y^{2}-4)^{3/2}\log\left[\frac{1}{2}\left(y^{2}-2-y\sqrt{y^{2}-4}\right)\right]\right].
\end{equation} 
Furthermore, the above relation holds in the limit where the $\rho$ parameter does not receive tree level contributions. This is a realistic scenario within our framework since, in the limit of vanishing triplet $vev$, the $\rho$ parameter does not deviate from unity at tree-level~\cite{9906332}. Thus, within the scenario of vanishing triplet $vev$, the scalar sector is parametrized by only three additional, independent parameters ($M_{\eta},a_{2},b_{4}$), since the mass of the SM Higgs boson is fixed at $125$ GeV. 

An extended Higgs sector containing multiplets in addition to the SM Higgs doublet can modify the Higgs couplings to fermions and gauge bosons. However, since an inert real scalar triplet does not mix with the SM Higgs doublet, tree-level modifications to the model's couplings do not exist. In particular, the couplings involving the scalar bosons are given by~\cite{1303.4490}
\begin{eqnarray} h^{0}\bar{f}f&:&-i\frac{m_f}{v},\quad
ZZh^{0}:\frac{2iM_Z^2}{v}g^{\mu\nu}, \quad
 \eta^+ \eta^-h^{0}:-ia_2v,\quad
W^+W^-h^{0}:ig^2\frac{1}{2}vg^{\mu\nu}, \nonumber\\
W^{+}&&\eta^-\eta^{0}:\frac{1}{2}(p'-p)^{\mu }\quad
\gamma \eta^+ \eta^-:ie\,\big(p'-p\big)^\mu,\quad
Z \eta^+ \eta^-: ig c_W\big(p'-p\big)^\mu,
\label{eq:coupling}\end{eqnarray}
where $g$ is the $SU(2)_{W}$ gauge coupling. In the case of an inert real triplet extension of the SM, the absence of mixing with the SM-like Higgs doublet results in the absence of couplings between $\eta^{0}$/$\eta^\pm$ and fermions, and therefore no additional contributions to the $Zb\bar{b}$ vertex are present~\cite{9906332}. However, since the triplet couples to electroweak gauge bosons at tree-level, it will generate one loop corrections to the gauge boson propagators, and thus contribute to the oblique parameters ($S$, $T$, $U$).

\subsection{Real Scalar Triplet with a Vector-like Electroweak Singlet Quark}\label{sec:tripletvector}

Vector-like quarks are an area of focus for LHC research, as colored objects are highly visible due to large cross sections at hadron colliders and they can affect the Higgs boson diphoton measurement through loop contributions to the effective vertex. Vector-like quarks are constrained both through effects in the flavor sector~\cite{AguilarSaavedra:2002kr,Botella:2012ju,Fajfer:2013wca,Aguilar-Saavedra:2013wba,Botella:2013bsa}, and through direct detection measurements~\cite{cmsHT,atlasHT,Tevatron}. 

In the previous section, the scalar triplet did not mix with the SM Higgs doublet, and therefore it had no Yukawa interactions with leptons and quarks. In this section, the scalar triplet couples to fermions and vector-like quarks through new non-renormalizable interactions parametrizing new physics at a scale $\Lambda~\sim 1$ TeV. This class of models is strongly motivated by the Little Higgs frameworks, where the SM-like Higgs boson is a pseudo-Goldstone boson of a large global symmetry explicitly broken by gauge, Yukawa, and scalar interactions.~\cite{ArkaniHamed:2002qy,Perelstein:2003wd,Schmaltz:2010ac,Martin:2013fta}. 

Recently, a number of studies have looked at models where additional scalars and vector-like quarks are introduced~\cite{Berger:2012ec,Wang:2012gm,Kearney:2013cca,Kearney:2013oia,Fukano:2013aea,Gillioz:2013pba,Xiao:2014kba,Karabacak:2014nca,Bahrami:2014ska}. Within the context of a vector-like $SU(2)_{W}$ singlet fermion, these studies have focused either on renormalizable interactions between the new scalar sector and the new fermion sector~\cite{Xiao:2014kba,Karabacak:2014nca}, or focused on renormalizable interactions induced through mixing that arises in the scalar sector and its effects on the SM Yukawa interactions~\cite{Wang:2012gm,Kearney:2013cca,Bahrami:2014ska}. Our approach is to introduce new non-renormalizable interactions between the new scalar sector and fermion sector in a scenario that allows mixing between the SM third generation up-type quark and the vector-like quark in a way that results in the cancellation of the leading quadratic divergences to the one-loop corrections to the mass of the Higgs boson.

We expand our toy model by extending the Yukawa sector of the Standard Model through the following dimension five operators:
\begin{eqnarray}
{\cal L}_{\mathrm{Yukawa}}&=&\bar Q (y_{1}+\epsilon_1 \frac{\Sigma}{\Lambda}) \tilde H u_R +  \bar Q (y_{2}+\epsilon_2 \frac{\Sigma}{\Lambda}) \tilde H \chi_R +  \bar Q (y_{b}+\epsilon_b\frac{\Sigma}{\Lambda}) H d_R \label{Ly} \nonumber \\
&+&\frac{y_3}{2\Lambda}H^\dagger H \bar\chi_L \chi_R+y_4 \Lambda \bar\chi_L \chi_R+\frac{y_5}{2\Lambda}H^\dagger H \bar\chi_L u_R + h.c.\label{eq:LagToy},
\end{eqnarray}
where $\bar Q=(\bar u_L,\bar d_L)$, $\tilde H=-i\sigma_2 H^{\star}$. We neglect interactions with the lighter generations of fermions. The effects of mixing between a single vector-like quark and all three generations of SM quarks have been recently studied in~\cite{Fajfer:2013wca}, including non-renormalizable interaction between quarks and the Higgs boson. Their study focuses on both di-Higgs and single Higgs couplings to quarks and takes into account all constraints arising from low energy flavor observables~\cite{Blankenburg:2012ex,Harnik:2012pb}. They show that significant modifications to these Higgs properties are possible and set bounds on the off-diagonal couplings between the heavy vector-like quark and the light generations. Within our study a similar approach could be taken, including a similar generalization of the $\epsilon_{i}$ couplings over all generations to include off-diagonal couplings to the light generations in the mass eigenstate basis. However, the off-diagonal couplings will be small since they would be modified by the mixing between the heavy vector-like quark with the top quark and the CKM terms involving the top quark and the light generations. In addition, a renormalizable term proportional to $\bar{\chi}_{L}u_{R}$ is not included, as it can be rotated away through a trivial field redefinition. We have ignored dimension five operators of the form $Tr[\Sigma^{2}]\bar{\chi}_{L}\chi_{R}$ since in the limit of small mixing between $H$ and $\Sigma$, the contributions from these operators to exotic decays of the heavy vector-like quark are negligible. The parameters $\epsilon_{1,2}$ are free parameters taken to be of order $\mathcal{O}(1)$.

We assume that the triplet scalar contributes negligibly to electroweak symmetry breaking (EWSB), and so the third generation up-type quarks, $u_{L,R}$, mix with the vector-like quarks, $\chi_{L,R}$ as in minimal $SU(2)_{W}$ singlet vector-like extensions of the SM~\cite{GenericSVQ}. The mass matrix between the third generation up-type quark and the heavy vector-like quark is given by
\begin{equation}
M_{T}=\begin{pmatrix}
\frac{y_{1} v}{\sqrt{2}} & \frac{y_{2} v}{\sqrt{2}} \\
\frac{y_{5} v^{2}}{4\Lambda} & y_4 \Lambda + y_3\frac{v^2}{4\lambda}
\end{pmatrix},\label{eq:TopMat}
\end{equation}
where $v$ is the $vev$ of the Higgs doublet. The mixing between the electroweak eigenstates can be parametrized in the following way:
\begin{equation}
 \left(\begin{array}{c} u_{L,R} \\ \chi_{L,R} \end{array}\right)=\left(\begin{array}{cc} c_{L,R} & s_{L,R} \\ -s_{L,R} & c_{L,R}\end{array}\right)  \left(\begin{array}{c} t_{L,R} \\ T_{L,R} \end{array}\right),
\end{equation}
where $s_{L,R}\equiv\sin\theta_{L,R}$ and $c_{L,R}\equiv\cos\theta_{L,R}$. These mixing angles can be expressed in terms of the parameters introduced in Equation~(\ref{eq:LagToy}) and expanded in inverse powers of $\Lambda$. To order $\Lambda^{-2}$, the mixing angles, in terms of the fundamental model parameters, are given by
\begin{eqnarray}
c_{L}&\approx&-1+\frac{y^{2}_{2}v^{2}}{4y^{2}_{4}\Lambda^{2}}, \qquad  s_{L}\approx\frac{y_{2}v}{\sqrt{2}y_{4}\Lambda}, \nonumber \\
c_{R}&\approx&1-{\cal O}\left(1/\Lambda^{4}\right), \qquad s_{R}\approx\frac{(2y_{1}y_{2}+y_{4}y_{5})}{4y^{2}_{4}\Lambda^{2}},\label{eq:mixang}
\end{eqnarray}
and the masses of the SM top quark and the heavy third generation up-type quark, $T$, are given by
\begin{eqnarray}
m^{2}_{t}&\approx&\frac{y^{2}_{1}v^{2}}{2}\left(1-\frac{v^2 y_{2}(y_{1}y_{2}+y_{4}y_{5})}{2y_1 y^{2}_{4}\Lambda^{2}}\right), \nonumber \\
m^{2}_{T}&\approx&y^{2}_{4}\Lambda^{2}\left(1+\frac{v^2(y_2^2+y_3 y_4)}{2 y_4^2 \Lambda^2}\right).\label{eq:masses}
\end{eqnarray}
Higher order terms in the expansion are taken into account in our numerical routines, in order to maintain consistency with powers of $v/\Lambda$.

Since we neglect the $vev$ of the triplet as small, one can use the general parametrization of the Lagrangian introduced in Equation~(\ref{eq:LagToy}), to express the couplings of the top quark and the heavy vector-like quark to the SM Higgs boson, $h^{0}$, as in~\cite{Fajfer:2013wca}
\begin{equation}
{\cal L}_{h^{0}}=\sum_{i,j}\left(-y_{ij}h^{0}+x_{ij}\frac{(h^{0})^{2}}{2v^{2}}\right)\bar{f}^{i}_{L}f^{j}_{R},
\end{equation}
where the sum is over $i,j=t,T$. The above parametrization can be used to express the condition for the cancellation of the quadratic divergences to the mass of the SM-like Higgs boson by
\begin{equation}
\sum_{i}x_{ii}\frac{m_{i}}{v}=\sum_{i,j}|y_{i,j}|^{2}.
\end{equation}
In terms of our toy model, this relationship can be expressed as~\cite{Fajfer:2013wca},
\begin{equation}
\frac{m^{2}_{t}c^{2}_{L}+m^{2}_{T}s^{2}_{L}}{v^{2}}\approx\frac{1}{\Lambda}\left[m_{t}s_{L}(-y_{5}c_{R}+y_{3}s_{R})+m_{T}c_{L}(y_{5}s_{R}+y_{3}c_{R})\right]\label{eq:QuadDiv},
\end{equation}
which is used to reduce the number of degrees of freedom in the quark sector by one.

This setup opens the possibility for new decay modes of the heavy top mass eigenstate, in particular, $T\to \eta^{0}t$ and $T\to \eta^{+}b$, in addition to the ones normally studied in minimal vector-like extensions of the SM ($T\to W^{+}b, ~th^{0},~tZ$). The relevant couplings involving these new modes are given by
\begin{eqnarray}
g_{\eta^{0}T\bar{t}}&=&\frac{v}{2\sqrt{2}\Lambda}((c_R s_L \epsilon_1 - s_R s_L \epsilon_2)P_L - (c_R c_L \epsilon_2 + s_R c_L \epsilon_1)P_R),\cr\cr
g_{\eta^{-}T\bar{b}}&=& \frac{v}{2\sqrt{2}\Lambda} ((s_R \epsilon_1+c_R \epsilon_2)P_R + s_L \epsilon_b P_L).
\label{eq:NewDecModes1}
\end{eqnarray}
Furthermore, because of the nature of the operators inducing these decays, the branching ratios to these new modes can be large in the small mixing region between the SM top quark and the heavy vector-like top quark. The new neutral scalar state then decays to $t\bar{t}^{(*)}$ and/or $b\bar{b}$, while the charged scalar decays to $t^{(*)}\bar{b}$, depending on the mass. The relevant couplings between the new scalar states and the $t$ and $b$ fermions are:
\begin{eqnarray}
g_{\eta^{0}t\bar{t}}&=&\frac{v}{2\sqrt{2}\Lambda}c_L(c_R \epsilon_1 - s_R \epsilon_2)(P_L - P_R),\cr\cr
g_{\eta^{0}b\bar{b}}&=&\frac{v}{2\sqrt{2}\Lambda}\epsilon_b(P_L-P_R),\cr\cr
g_{\eta^{-}t\bar{b}}&=& \frac{v}{2\sqrt{2}\Lambda}((c_R \epsilon_1 - s_R \epsilon_2)P_R + c_L \epsilon_b P_L).
\label{eq:NewDecModes2}
\end{eqnarray}

Given the constraints on the top mass (Equation~(\ref{eq:masses})), $m_t = 173$ GeV, and the cancellation of the quadratic divergences, Equation~(\ref{eq:QuadDiv}), we reduce our degrees of freedom in the heavy quark sector by two. Furthermore, we choose the more phenomenological parameters of the heavy top mass, $m_T$, and the sine of the left-handed mixing angle, $s_L$, leaving $\Lambda$, $y_5$, $\epsilon_1$ and $\epsilon_2$ as the remaining fundamental parameters. These remaining parameters we fix for several different scenarios and use Equations~(\ref{eq:mixang}),(\ref{eq:masses}), and (\ref{eq:QuadDiv}) to solve for $y_{1}-y_{4}$. In addition, since the bare mass of the vector-like quarks is given by $y_{4}\cdot\Lambda$, the validity of the effective model will be for values of $y_{4}\lesssim1$. Our results are shown in the $s_L - m_T$ plane.

At the one-loop level, the dimension five operators will necessarily generate an $a_1$ term, which was previously neglected, in addition to contributing to the other scalar parameters. Due to the lack of constraints on the other parameters, we are free to absorb the one-loop contributions to the other scalar parameters without loss of generality. Ignoring the logarithmic contribution as sub-leading, we find:
\begin{equation}
a_1^{1loop} = \frac{(y_1 \epsilon_1 + y_2 \epsilon_2)\Lambda}{32\pi^2}
\end{equation}
Thus, maintaining the limit of $a_1^\prime = a_1 + a_1^{1loop} \rightarrow 0$ represents a large degree of fine-tuning. To avoid fine tuning, we must assume that $a_1^\prime$ is not significantly smaller than the largest of $a_1$ or $a_1^{1loop}$, and re-consider the constraints that come from the triplet $vev$. As before, taking $b_4$ and $a_2$ to be $\mathcal{O}(1)$, with $M_\Sigma$ to be $\mathcal{O}(100)$~GeV, the constraint on $\delta \rho = (2v_3/v_0)^2 \leq 0.001$ translates to a constraint on $a_1$ to be $\mathcal{O}(10)$~GeV. Except for very small values of $s_L$, we have ensured that this constraint is not violated over the entire parameter space that we consider. In addition, our assumption that the triplet $vev$ does not contribute significantly to the masses of the fermions is an acceptable approximation, since $v_3/v_0 \ll 1$.

\section{Constraints}\label{sec:constraints}

Constraints on our model come from three primary sources - contributions to the oblique parameters ($S$, $T$, $U$), extra one-loop contributions to the $Zb\bar{b}$ vertex, and direct collider constraints from searches for heavy vector-like quarks.

\subsection{Oblique Parameters}
The corrections to $S$, $T$ and $U$ can be parametrized as 
\begin{eqnarray}
\alpha S&=&\frac{4s^{2}_{W}c^{2}_{W}}{M_{Z}}\left(\Delta\Pi^{ZZ}(M_{Z})-\frac{c^{2}_{W}-s^{2}_{W}}{s_{W}c_{W}}\Delta\Pi^{\gamma Z}(M_{Z})-\Delta\Pi^{\gamma\gamma}(M_{Z})\right), \nonumber \\
\alpha T&=&\frac{1}{M^{2}_{W}}\left(\Pi^{WW}(0)-c^{2}_{W}\Pi^{ZZ}(0)\right), \nonumber \\
\alpha\left( S+U \right)&=&4s^{2}_{W}\left(\frac{\Delta\Pi^{WW}(M_{W})}{M^{2}_{W}}-\frac{c_{W}}{s_{W}}\frac{\Delta\Pi^{\gamma Z}(M_{Z})}{M^{2}_{Z}}-\frac{\Delta\Pi^{\gamma\gamma}(M_{Z})}{M^{2}_{Z}}\right),
\end{eqnarray}
where $\Delta\Pi(k)=\Pi(k)-\Pi(0)$, the functions $\Pi(k)$ denote the coefficients of the metric in the gauge boson inverse propagators, $\alpha$ is the fine structure constant and $c_{W},s_{W}$ are the cosine and sine of the Weinberg angle respectively. The current experimental bounds on the oblique parameters are~\cite{Baak:2012kk}
\begin{eqnarray}
\Delta T&=&T-T_{SM}=0.08\pm0.07, \nonumber \\
\Delta S&=&S-S_{SM}=0.05\pm0.09, \nonumber \\
\Delta U&=&U-U_{SM}=0. \nonumber 
\end{eqnarray}

Contributions to the oblique parameters from a real scalar triplet have been studied in~\cite{Forshaw:2001xq,Forshaw:2003kh,Chen:2006pb,Chankowski:2006hs,SekharChivukula:2007gi,0811.3957}, and are given by
\begin{eqnarray}
S_{TM}&=&0, \nonumber \\
T_{TM}&\approx&\frac{1}{6\pi}\frac{1}{s^{2}_{W}c^{2}_{W}}\frac{\Delta M^{2}}{M^{2}_{Z}}, \nonumber \\
U_{TM}&\approx&\frac{\Delta M}{3\pi M_{\eta^\pm}},
\end{eqnarray}
in the limit of small $\Delta M$, where $\Delta M\equiv M_{\eta^{0}}-M_{\eta^\pm}$. In the limit of vanishing triplet $vev$ and no couplings to fermions, the contributions to the $T$ and $U$ parameters are largely suppressed, since the mass difference between the charged and neutral components of $\Sigma$ only arise due to radiative corrections coming from the coupling of $\eta^\pm$ to the $Z$ and $W$ gauge bosons. The additional contribution to the mass splitting from the couplings to the heavy quark sector is also expected to be small, as all couplings are further suppressed by factors of $v/\Lambda$.

Corrections to the oblique parameters from the heavy quark sector arise solely due to the mixing between $u_{L,R}$ and $\chi_{L,R}$. In particular, only one-loop corrections to the $S$ and $T$ parameters arise. These are given by~\cite{Xiao:2014kba}
\begin{eqnarray}
\Delta T_{T}&=&T^{SM}_{t}s^{2}_{L}\left[-(1+c^{2}_{L})+s^{2}_{L}\frac{m^{2}_{T}}{m^{2}_{t}}+c^{2}_{L}\frac{2m^{2}_{T}}{m^{2}_{T}-m^{2}_{t}}\log \frac{m^{2}_{T}}{m^{2}_{t}}\right], \nonumber \\
\Delta S_{T}&=&-\frac{s^{2}_{L}}{6\pi}\left[(1-3c^{2}_{L})\log\frac{m^{2}_{T}}{m^{2}_{t}}+5c^{2}_{L}-\frac{6c^{2}_{L}m^{4}_{t}}{(m^{2}_{T}-m^{2}_{t})^{2}}\left(\frac{2m^{2}_{T}}{m^{2}_{t}}-\frac{3m^{2}_{T}-m^{2}_{t}}{m^{2}_{T}-m^{2}_{t}}\log\frac{m^{2}_{T}}{m^{2}_{t}}\right)\right], \nonumber \\
\nonumber \\
\end{eqnarray}
where 
\begin{equation}
T^{SM}_{t}=\frac{3m^{2}_{t}}{16\pi s^{2}_{W}}\frac{m^{2}_{t}}{M^{2}_{W}},
\end{equation}
denotes the SM contribution to the $T$ parameter that arises from a loop of SM top and bottom quarks. From the above two equations one can easily see that this constraint is strong in the large mixing limit of our model. In particular, these constraints are identical to those that arise within a simple renormalizable extension of the SM Yukawa sector with a pair of $SU(2)_{W}$ singlet vector-like quarks, $\chi_{L,R}$~\cite{GenericSVQ}. Within this class of models, a $400$ GeV heavy top quark is ruled out in the region where $s_{L}\gtrsim0.2$ and the constraint on $s_{L}$ becomes stronger for larger values of the heavy top mass, $m_{T}$. Therefore, we expect our toy model to be restricted to within the region of parameter space with small $s_L$.

\subsection{$Z\rightarrow b\bar{b}$}

The effective $Zb\bar{b}$ coupling has been measured with excellent accuracy at LEP
and forms a strong constraint on new physics. Within the SM, the $Zb\bar{b}$ vertex, including leading one-loop contributions from the top quark, can be parametrized by the following couplings:
\begin{eqnarray}
g^{SM}_{L}&=&-\frac{1}{2}+\frac{1}{3}s^{2}_{W}+\frac{m^{2}_{t}}{16\pi^{2}v^{2}}, \nonumber \\
g^{SM}_{R}&=&\frac{1}{3}s^{2}_{W},\label{eq:effcoup}
\end{eqnarray}
where the above expressions have been normalized by a factor of $g/\sqrt{1-s^{2}_{W}}$. Within this toy model, contributions from both the mixing between the SM top quark and the heavy fermion, as well as the tree-level coupling of the charged scalar to the $Z$ gauge boson given in Equation~(\ref{eq:coupling}) lead to deviations from the SM predictions of the following precision observables on the $Z$ resonance~\cite{Baak:2012kk}:
\begin{eqnarray}
R^{SM}_{b}&=&0.21474\pm0.00003, \nonumber \\
A^{SM}_{b,FB}&=&0.1032^{+0.0004}_{-0.0006}, \nonumber \\
A^{SM}_{b}&=&0.93464^{+0.00004}_{-0.00007}, \nonumber \\
R^{SM}_{c}&=&0.17223\pm0.00006,
\end{eqnarray}
where $R^{SM}_{b,c}$ denote the fraction of $b$- and $c$-quarks produced in $Z$ decays and $A^{b,SM}_{FB}$ \& $A^{SM}_{b}$ denote the forward-backward and polarized asymmetries, respectively, in the production of $b$-quarks from $Z$ decays as predicted by the SM. Using the first order expressions found in~\cite{GenericSVQ}, any deviation from the SM prediction may be factorized as:
\begin{eqnarray}
R_{b}&=&R^{SM}_{b}\left(1-1.820\delta g_{L}+0.336\delta g_{R}\right), \nonumber \\
A^{b}_{FB}&=&A^{b,SM}_{FB}\left(1-0.1640\delta g_{L}-0.8877\delta g_{R}\right), \nonumber \\
A_{b}&=&A^{SM}_{b}\left(1-0.1640\delta g_{L}-0.8877\delta g_{R}\right), \nonumber \\
R_{c}&=&R^{SM}_{c}\left(1+0.500\delta g_{L}-0.0924\delta g_{R}\right),
\end{eqnarray}
where $\delta g_{L}$ and $\delta g_{R}$ denote the shifts in the effective coupling introduced in Equation~(\ref{eq:effcoup}). 

\begin{figure}[!h]
\begin{center}
\includegraphics[width=0.7 \textwidth]{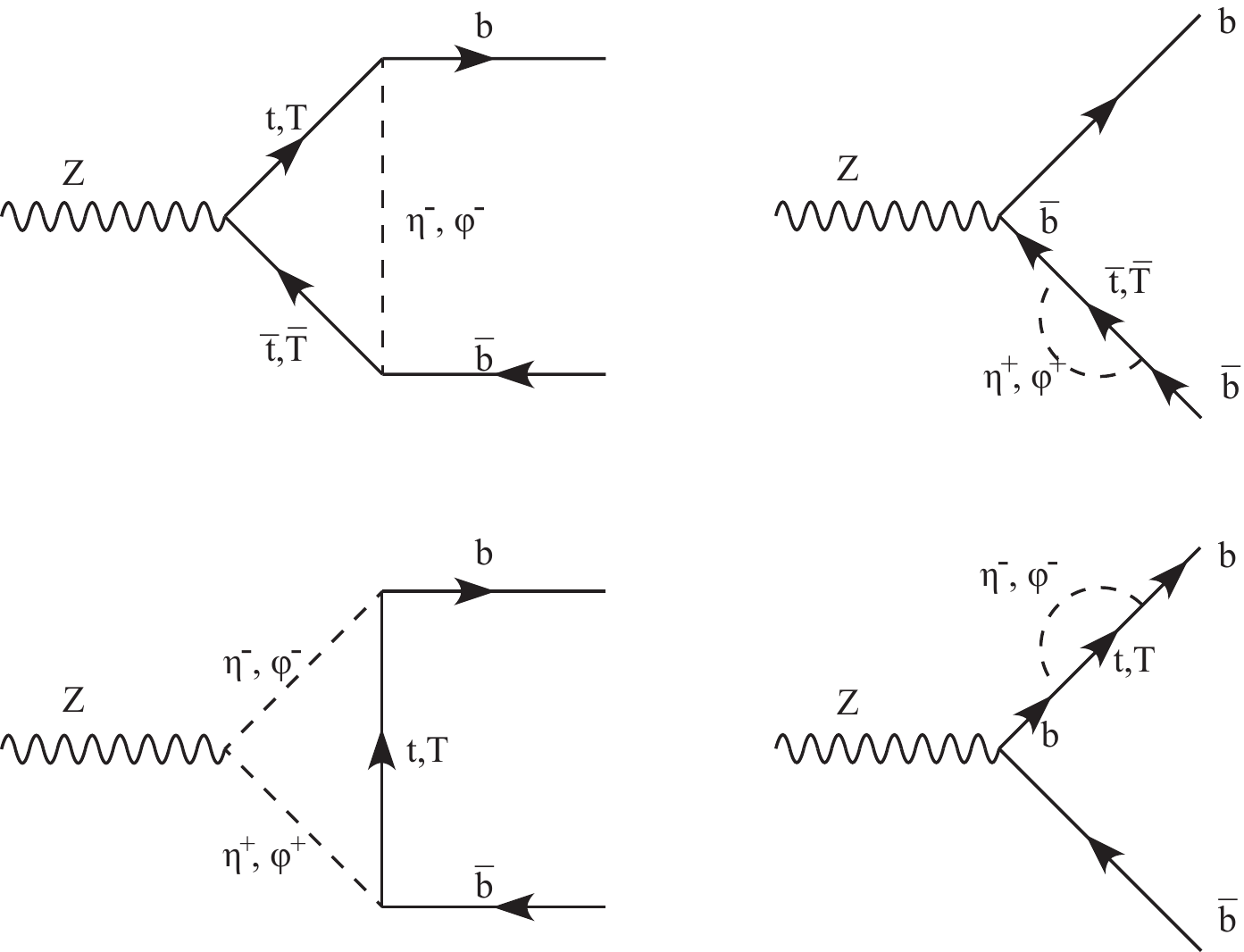}
\caption{Feynman diagrams of new contributions to the effective $Z \rightarrow b\bar{b}$ vertex.}
\label{fig:diagrams}
\end{center}
\end{figure}

In the 't Hooft-Feynman gauge, one-loop corrections to $\delta g_{L}$ arise from loops where the longitudinal components of the $W$ and $Z$ gauge bosons are just the Goldstone modes, $\phi^{\pm}$ and $Im(\phi^{0})$ in Equation~(\ref{eq:ScFields}), and when accounting for mixing between the heavy top quark, $T$, and the SM top quark, $t$. Additional one-loop contributions also arise from the new charged scalar, $\eta^\pm$. The new diagrams are summarized in Figure~\ref{fig:diagrams}.
The leading contributions to $\delta g_{L}$ from the Goldstone modes, including the mixing between $t$ and $T$, are proportional to $y_{1}$ and $y_{2}$ and can be expressed as
\begin{eqnarray}
\delta g_{L}[\phi^{\pm}]&=&\frac{\sqrt{1-s^{2}_{W}}}{16\pi^{2}g}\left[-(g^{\phi^- t\bar{b}}_{L})^{2}\left(-2g^{Zt\bar{t}}_{R}C_{24}+\frac{1}{2}g^{Zt\bar{t}}_{R}+g^{Zt\bar{t}}_{L}m^{2}_{t}C_{0}\right)\right. \nonumber \\
&-&\left.(g^{\phi^- T\bar{b}}_{L})^{2}\left(-2g^{ZT\bar{T}}_{R}C_{24}+\frac{1}{2}g^{ZT\bar{T}}_{R}+g^{ZT\bar{T}}_{L}m^{2}_{T}C_{0}\right)\right. \nonumber \\
&-&\left. g^{\phi^- t\bar{b}}_{L}\cdot  g^{\phi^- T\bar{b}}_{L} \left(-2g^{Zt\bar{T}}_{R}C_{24}+\frac{1}{2}g^{Zt\bar{T}}_{R}+g^{Zt\bar{T}}_{L}m_{t}m_{T}C_{0}\right)\right],\label{eq:Zbb1}
\end{eqnarray}
while the leading contributions from the charged scalar, $\eta^\pm$, are proportional to $\epsilon_{1}$ and $\epsilon_{2}$ and are given by
\begin{eqnarray}
\delta g_{L}[\eta^\pm]&=&\frac{\sqrt{1-s^{2}_{W}}}{16\pi^{2}g}\left[-(g^{\eta^- t\bar{b}}_{L})^{2}\left(-2g^{Zt\bar{t}}_{R}C_{24}+\frac{1}{2}g^{Zt\bar{t}}_{R}+g^{Zt\bar{t}}_{L}m^{2}_{t}C_{0}\right)\right. \nonumber \\
&-&\left.(g^{\eta^- T\bar{b}}_{L})^{2}\left(-2g^{ZT\bar{T}}_{R}C_{24}+\frac{1}{2}g^{ZT\bar{T}}_{R}+g^{ZT\bar{T}}_{L}m^{2}_{T}C_{0}\right)\right. \nonumber \\
&-&\left. g^{\eta^- t\bar{b}}_{L}\cdot  g^{\eta^- T\bar{b}}_{L} \left(-2g^{Zt\bar{T}}_{R}C_{24}+\frac{1}{2}g^{Zt\bar{T}}_{R}+g^{Zt\bar{T}}_{L}m_{t}m_{T}C_{0}\right)\right].\label{eq:Zbb2}  
\end{eqnarray}
The three-point integral factors, $C_0$ and $C_{24}$ can be found in \cite{9906332,CapdequiPeyranere:1990qk}, where we have used the definitions:
\begin{eqnarray}
C_{0}&\equiv&C_{0}(m^{2}_{b},M^{2}_{Z},m^{2}_{b};m^{2}_{i},M^{2}_{S},m^{2}_{j}), \nonumber \\
C_{24}&\equiv&C_{24}(m^{2}_{b},M^{2}_{Z},m^{2}_{b};m^{2}_{i},M^{2}_{S},m^{2}_{j}),
\end{eqnarray}
where $m_{i,j}=m_{t},m_{T}$ and $M_{S}$ denotes the mass of either the charged goldstone mode (with mass equal to the mass of $W$ gauge boson), or of the charged scalar, $\eta^\pm$. The couplings between the charged scalars and fermions in Equations~(\ref{eq:Zbb1})-(\ref{eq:Zbb2}) are given by
\begin{eqnarray}
g^{\phi^- t\bar{b}}_{L}&=&-y_{1}c_{R}+y_{2}s_{R}, \nonumber \\
g^{\phi^- T\bar{b}}_{L}&=&-y_{1}s_{R}-y_{2}c_{R}, \nonumber \\
g^{\eta^-t\bar{b}}_{L}&=&\frac{v}{2\Lambda}\left(\epsilon_{1}c_{R}-\epsilon_{2}s_{R}\right), \nonumber \\
g^{\eta^-T\bar{b}}_{L}&=&\frac{v}{2\Lambda}\left(\epsilon_{1}s_{R}+\epsilon_{2}c_{R}\right),
\end{eqnarray}
while the couplings between the $Z$ gauge boson and fermions are given by
\begin{eqnarray}
g^{Zt\bar{t}}_{L}&=&g_{W}\left(\frac{c^{2}_{L}}{2}-\frac{2}{3}s^{2}_{W}\right), \nonumber \\
g^{Zt\bar{t}}_{R}&=&g_{W}\left(-\frac{2}{3}s^{2}_{W}\right), \nonumber \\
g^{ZT\bar{T}}_{L}&=&g_{W}\left(\frac{s^{2}_{L}}{2}-\frac{2}{3}s^{2}_{W}\right), \nonumber \\
g^{ZT\bar{T}}_{R}&=&g_{W}\left(-\frac{2}{3}s^{2}_{W}\right), \nonumber \\
g^{Zt\bar{T}}_{L}&=&g_{W}s_{L}c_{L}, \nonumber \\
g^{Zt\bar{T}}_{R}&=&0,
\end{eqnarray}
with $g_{W}\equiv g/\sqrt{1-s^{2}_{W}}$.

Our constraints from the $Z \rightarrow b\bar{b}$ measurements are based on the latest experimental results~\cite{ALEPH:2005ab}:
\begin{eqnarray}
R^{exp}_{b}&=&0.21629\pm0.00066, \nonumber \\
A^{exp,b}_{FB}&=&0.0992\pm0.0016, \nonumber \\
A^{exp}_{b}&=&0.923\pm0.020, \nonumber \\
R^{exp}_{c}&=&0.1721\pm0.003.
\end{eqnarray}
We calculate a $95\%$ confidence level upper limit on each individual observable assuming that the contributions to $\delta g_{R}$ are negligible, since these are proportional to the bottom Yukawa coupling, $y_{b}$, when the Goldstone mode propagates in the loop and proportional to $\epsilon_{b}$ for a charged scalar, $\eta^\pm$. We find that the strongest limit is set by $R_{b}$ which constrains the deviation on the $g_{L}$ in the range $-0.00568<\delta g_{L} <0.00298$. Furthermore, it has been shown that the measurement of ${\cal B}(B_{s}\to\mu^{+}\mu^{-})$ can be used to constrain new physics models that predict modifications to the $Zb\bar{b}$ vertex, in particular models with an underlying flavor structure for the new physics~\cite{Bsmumu}. However, the constraint on the $Zb\bar{b}$ vertex correction used in our analysis is comparable to that derived from ${\cal B}(B_{s}\to\mu^{+}\mu^{-})$. We find that the constraints arising from corrections to the oblique parameters place far more stringent limits on the parameter space of this model. 

\subsection{Searches for heavy, vector-like quarks at the LHC\label{sec:LHCTTbar}}

Both CMS~\cite{cmsHT} and ATLAS~\cite{atlasHT} have performed searches for heavy, vector-like, charge $+2/3$ quarks, assuming that these states can decay to only three possible final states, $T \rightarrow W^+ b$, $T \rightarrow tZ$ and $T \rightarrow th^{0}$, with the sum of the branching ratios equalling unity. With the masses of the decay products well known, a thorough analysis of the acceptance rates is determinable for all signal regions, and accurate lower limits can be extrapolated for any model with a heavy quark that is limited to these decay modes. However, these results are not immediately transferable to our toy model due to the possibility of extra decay modes.

The idea of using an existing analysis to constrain beyond the SM (BSM) scenarios and applying it to a different BSM scenario has been studied very recently and introduced as a {\it data recasting} procedure to set limits on extensions of the SM~\cite{Barducci:2014ila}. We perform a similar data recasting analysis, except accounting for the extra decay modes allowed in our toy model.

The analyses carried out by the ATLAS and CMS collaborations assume pair production of the heavy top quark. This production mode is dominated by QCD production, and the cross section is determinable in a model independent fashion from the work in~\cite{Berger:2009qy}, or using the HATHOR coding package~\cite{HATHOR}. In particular, we focus on the CMS results and similarly use the HATHOR package to calculate our production cross sections. The CMS study establishes four signal regions (SR) that are sensitive to the presence of new heavy quarks with masses above 500 GeV: opposite-sign dilepton with two or three jets (OS1), opposite sign dilepton with five or more jets (OS2), same-sign dilepton (SS), and trilepton (Tri). The branching ratio independent efficiencies have been provided on the CMS wiki page for the study, showing the acceptance efficiency for all six combinations of $tZ$, $Wb$ and $th^{0}$ branching ratios.

For each channel, $k$, the CMS study has provided the number of observed events $N^{obs}_k$, as well as the number of expected background events with a corresponding uncertainty. From these values, we have determined the 95\% C.L. excluded number of signal events, $N_k^{95}$, using the single-channel $CL_s$ method, adapted from the CHECKMate program~\cite{checkmate}. For $k=$ (OS1, OS2, SS, Tri), the values of $N_k^{95}$ are (12.05, 30.43, 13.16, 5.58), assuming a Gaussian distributed probability distribution function for the uncertainty on the background events, and a negligible uncertainty on the signal events.

\begin{figure}[!h]
\begin{center}
\includegraphics[width=0.7 \textwidth]{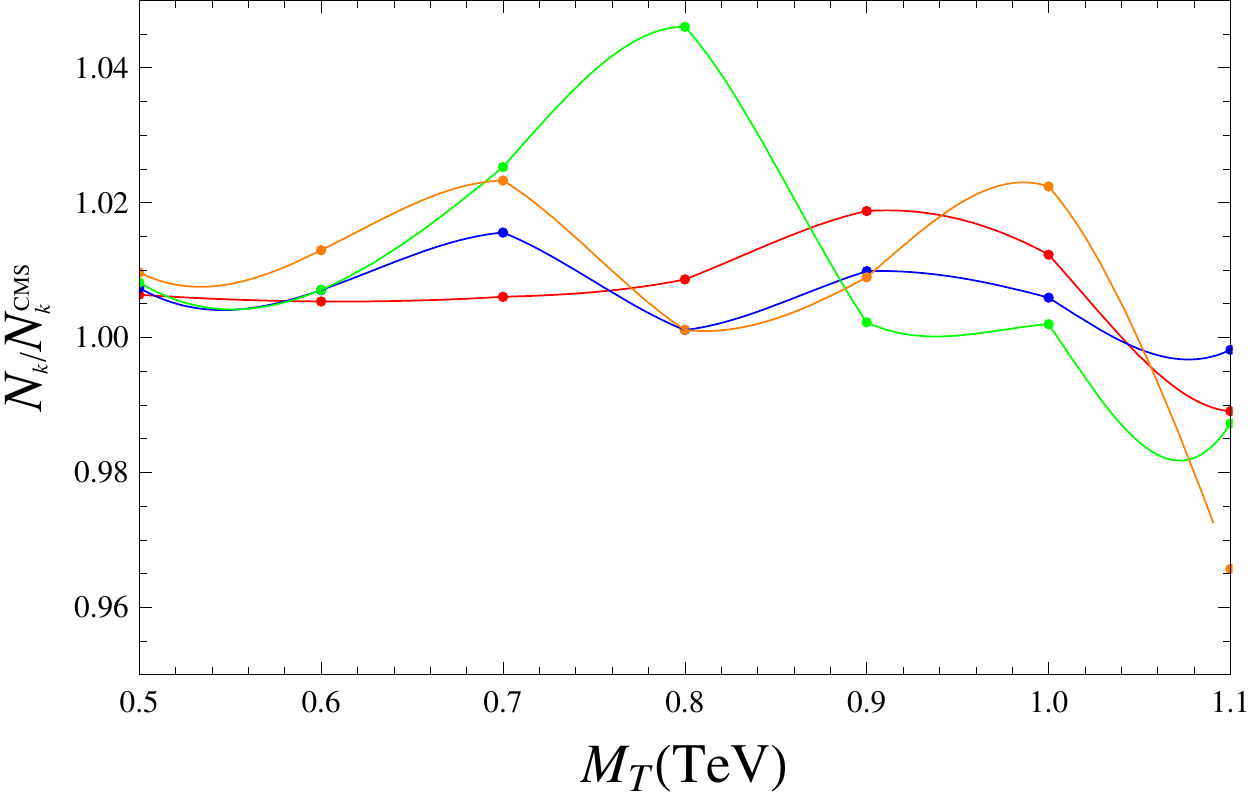}
\caption{Ratio of our calculated event rates to the CMS quotes event rates for the $(bW,tZ,th)=(0.5,0.25,0.25)$ branching ratio point. The red line corresponds to the OS1 signal region, blue to the OS2 signal region, green to the SS signal region, and orange to the Tri signal region.}
\label{fig:plot2}
\end{center}
\end{figure}

The acceptance efficiency, $\epsilon^k_i$, for each permutation, $i$, of two of the decay modes ($bW$, $tZ$ and $th^{0}$) is provided for each of the four signal regions, $k$, in the Wiki page for the CMS study.
From these, the number of signal events can be calculated as 
\begin{equation}
N_k(M_T)=\mathcal{L}\sigma_{T\bar{T}}(M_T)\displaystyle\sum_i \epsilon^k_i BR(T\bar{T} \rightarrow i),
\end{equation}
for integrated luminosity $\mathcal{L}$ and cross section $\sigma(T\bar{T})$ calculated with HATHOR. This is the identical procedure described in \cite{cmsHT}. The CMS study provides a list of the number of signal events they calculated for the $(bW,tZ,th^{0})=(0.50,0.25,0.25)$ branching ratio point, $N_k^{CMS}(M_T)$, which we use to compare our calculation. Figure \ref{fig:plot2} shows the comparison of our calculated signal events for $M_T$ between 500 and 1100 GeV, which amounts to at most a 4\% difference. This difference is due to the rough rounding in the quoted CMS results, which has a larger effect on the smaller event rates that occur at higher masses. However, these have a negligible effect on our final results. 

To estimate the acceptance rate for the new decay modes, we scale the provided acceptance efficiencies by the ratio of branching ratios that produce the tagged states for each of the signal regions. For the $t\eta^0$ final state, the following acceptance efficiencies were used:
\begin{equation}
\epsilon^k_{t\eta^0+i}(m_T) = \epsilon^k_{th+i}(m_T) \frac{BR(t\eta^0+i \rightarrow k)}{BR(th^{0}+i \rightarrow k)},
\end{equation}
where $k$ indicates the signal region (OS1, OS2, SS, Tri) as described previously, and $i$ represents the other decay mode ($bW$, $tZ$, $th^{0}$). Similarly for the $b\eta^\pm$ decay mode, the following acceptance efficiency was used:
\begin{equation}
\epsilon^k_{b\eta^\pm+i}(m_T) = \epsilon^k_{bW+i}(m_T) \frac{BR(b\eta^\pm+i \rightarrow k)}{BR(bW+i \rightarrow k)}.
\end{equation}
The results are relatively insensitive to changes in choice between $bW$ and $th^{0}$ for the charged $\eta$ decay mode, and between $tZ$ and $th^{0}$ for the neutral decay mode, indicating that the branching-ratio-independent acceptance rates (ratio of the efficiency to the branching ratio combinations that identically reproduces the signal region of interest) are more dependent on the masses and kinematics than the decay mode itself.

With this approach, we extend the CMS analysis and incorporate additional $T$-quark decay modes, $T\to X$, using the extracted efficiencies to set new limits on the vector-like top quark mass. We use the branching-ratio-independent acceptance rates, in combination with the branching ratios for the new decay modes (and the relevant branching ratios of the $\eta^0$ and $\eta^\pm$), to estimate the number of events for each SR. We assume that the new scalars decay exclusively to third generation quarks ($t\bar{t}$ and $t\bar{b}$), ignoring small CKM mixing effects. The new scalars are forbidden from decaying to pairs of gauge bosons, and the lack of mixing with the $h^{0}$ prevents the possibility of an $\eta \rightarrow V h^{0}$ decay mode.

\begin{figure}[!hp]
	\includegraphics[width=0.43\textwidth]{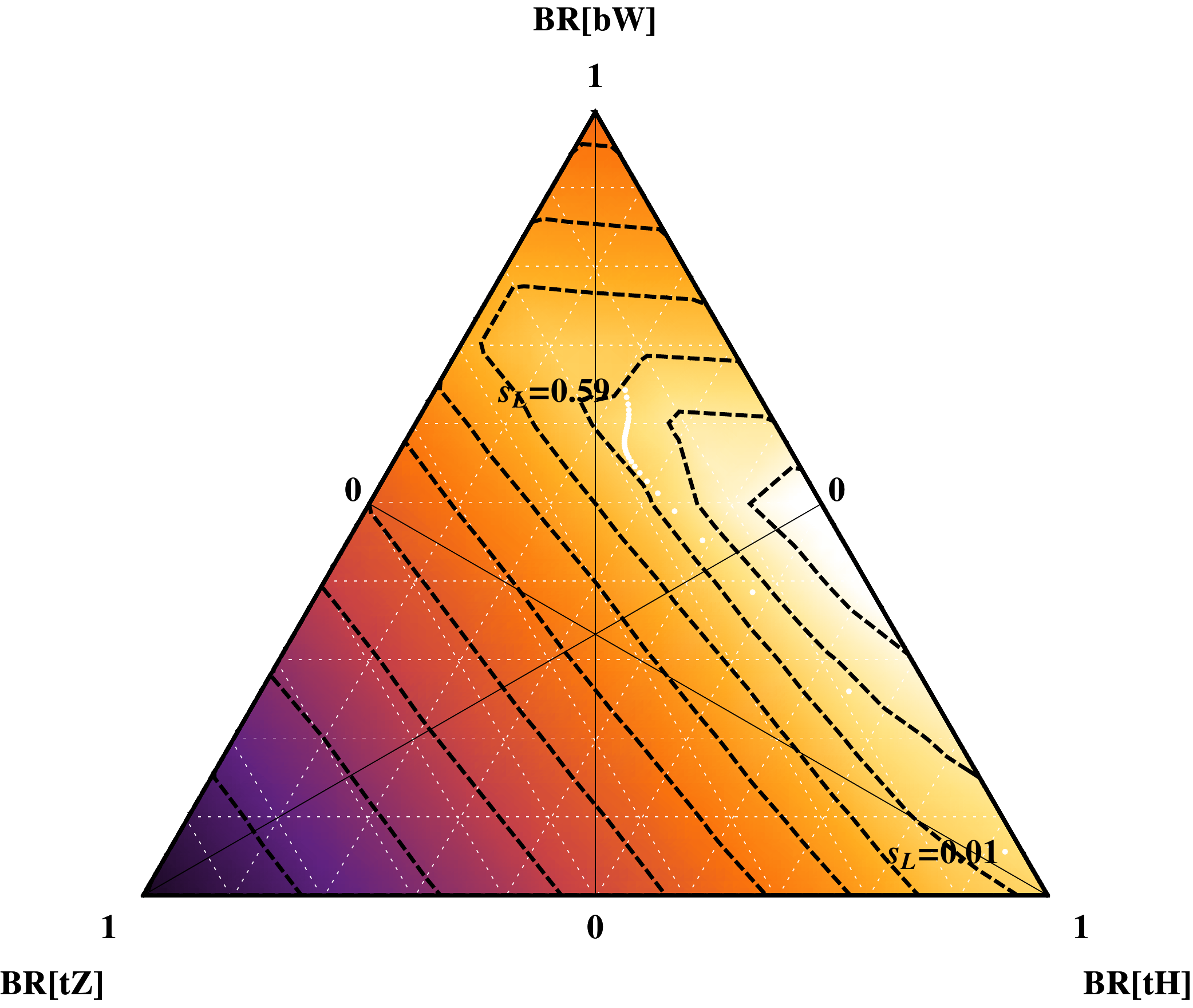}
	\includegraphics[width=0.43\textwidth]{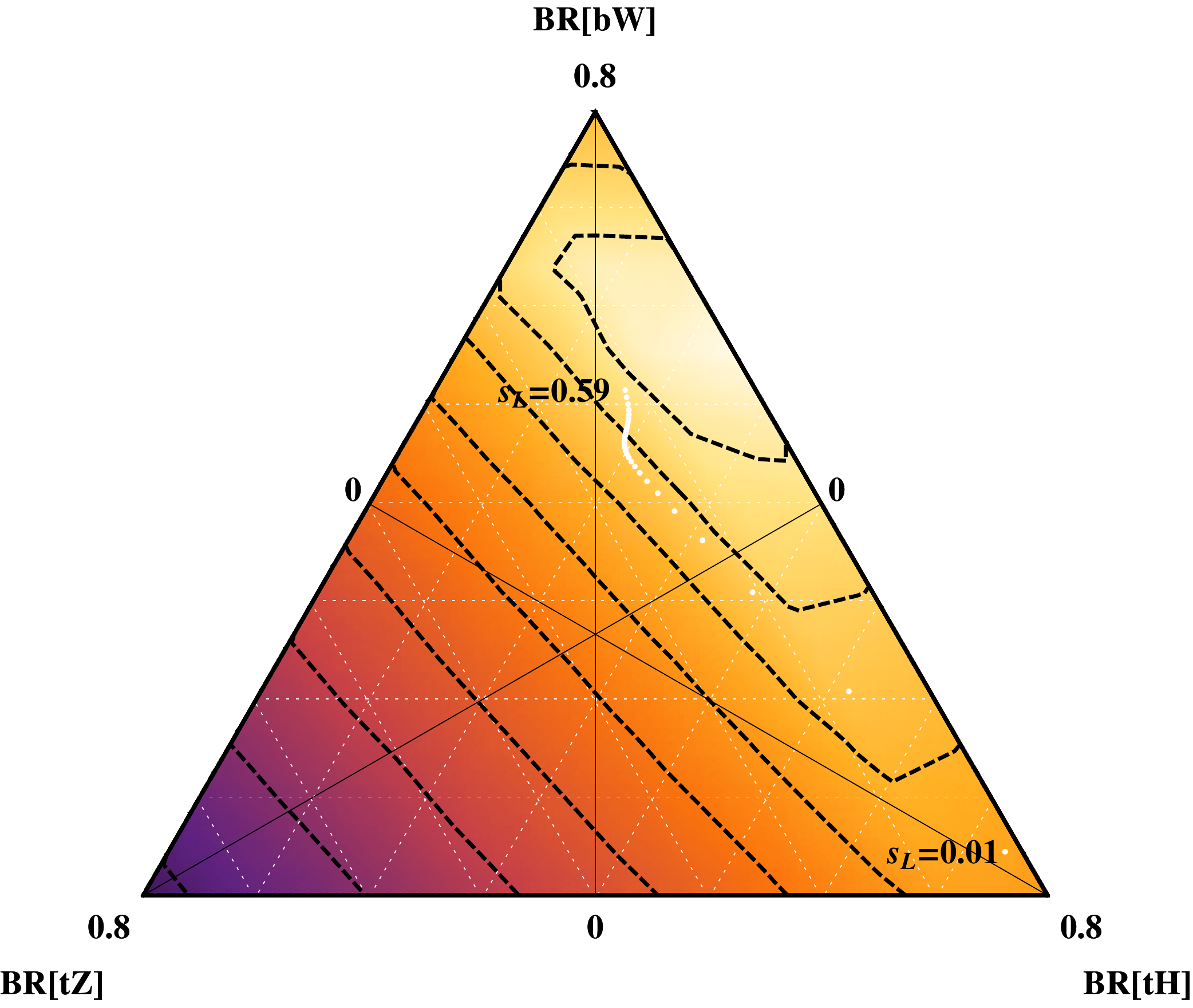}\\
	\includegraphics[width=0.43\textwidth]{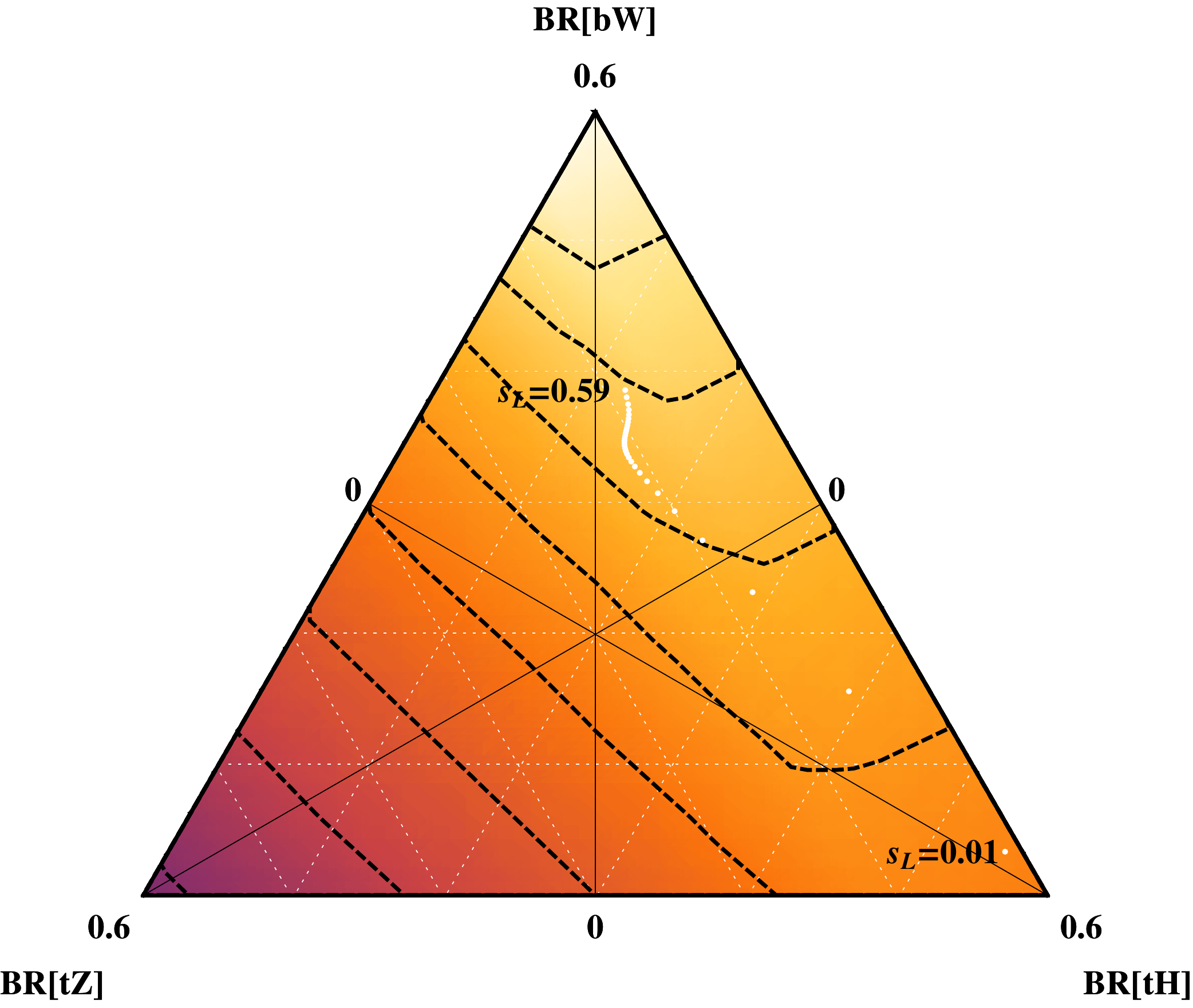}
	\includegraphics[width=0.43\textwidth]{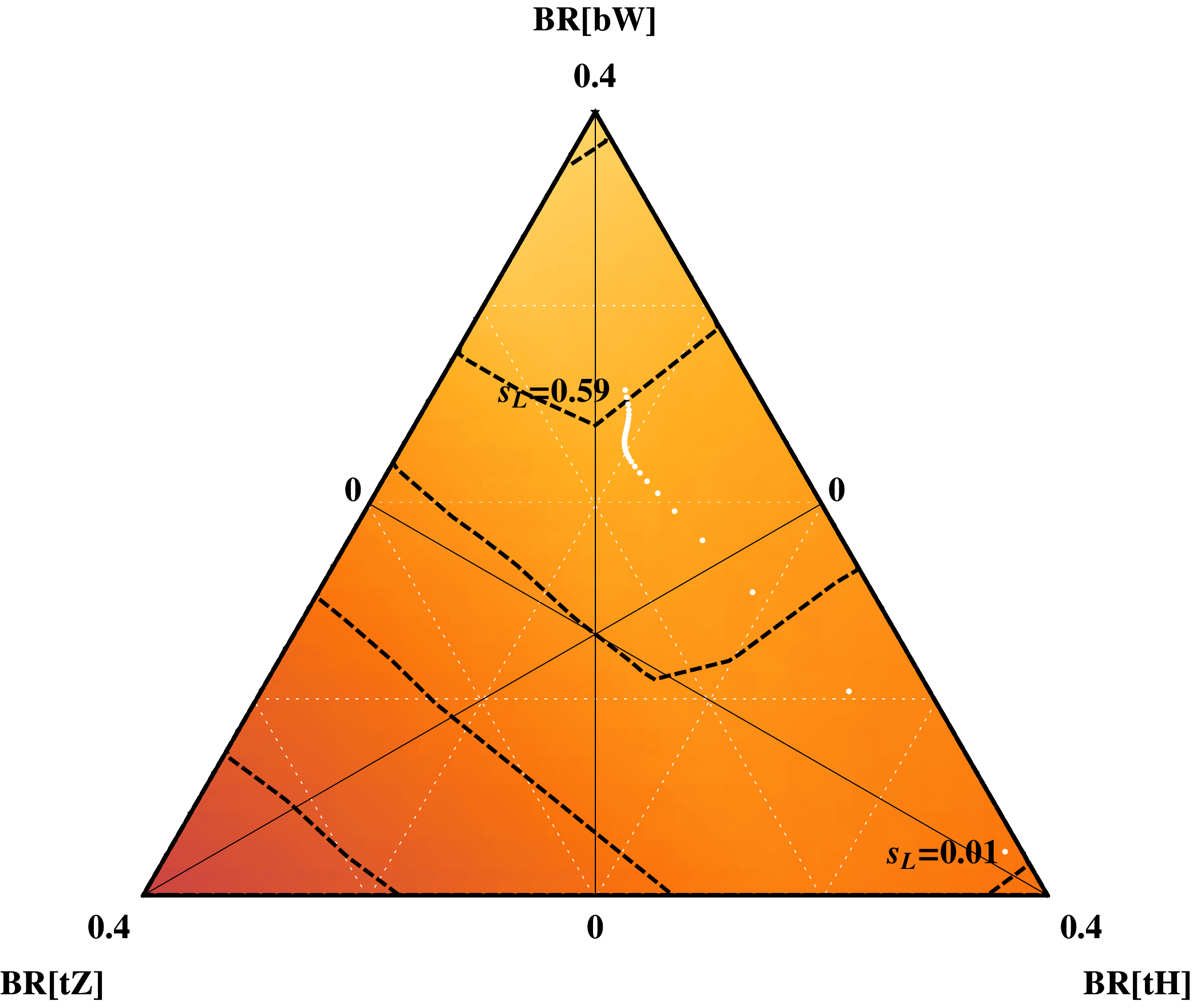}\\
	\includegraphics[width=0.43\textwidth]{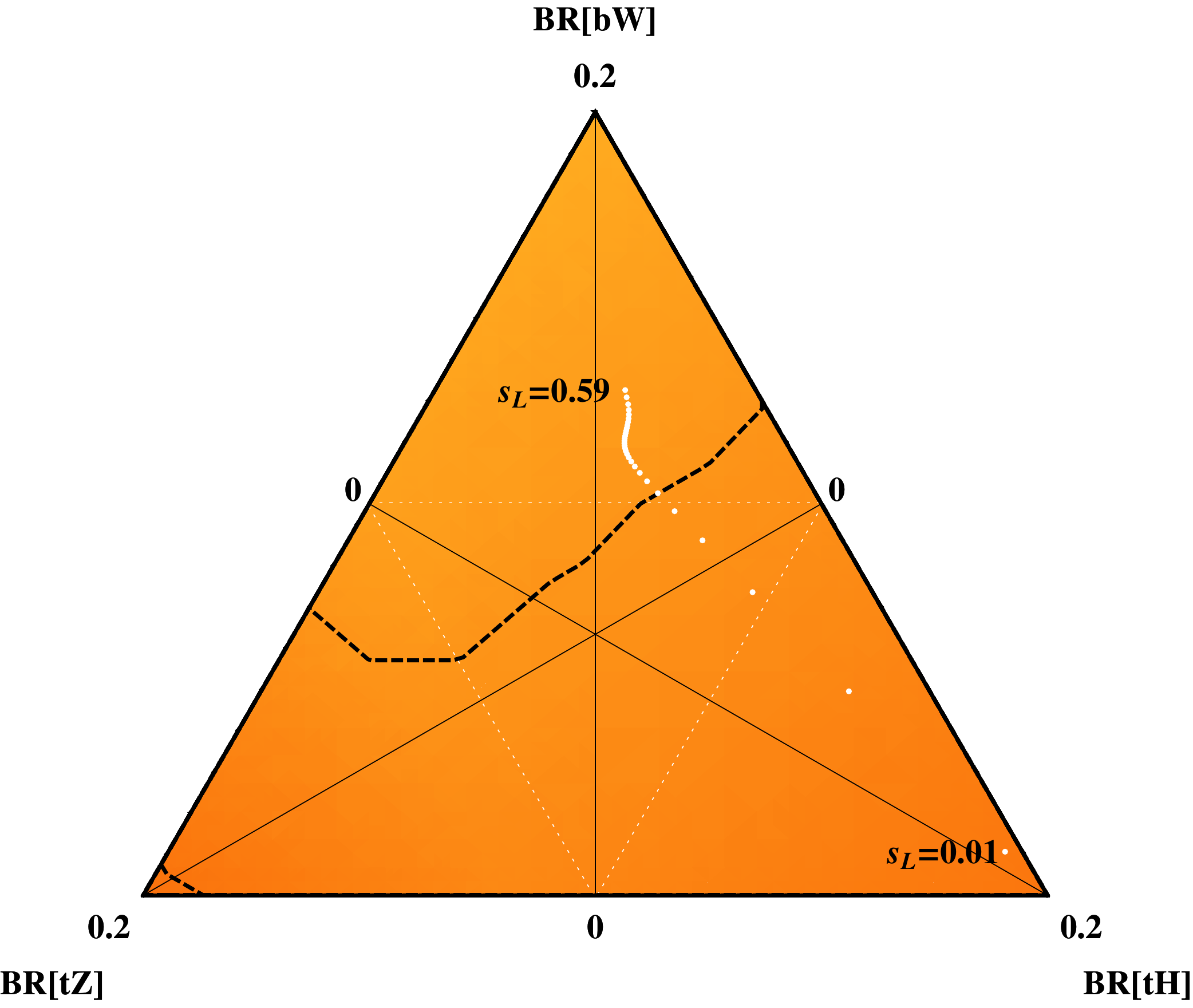}\hspace{28mm}
	\includegraphics[width=0.08\textwidth]{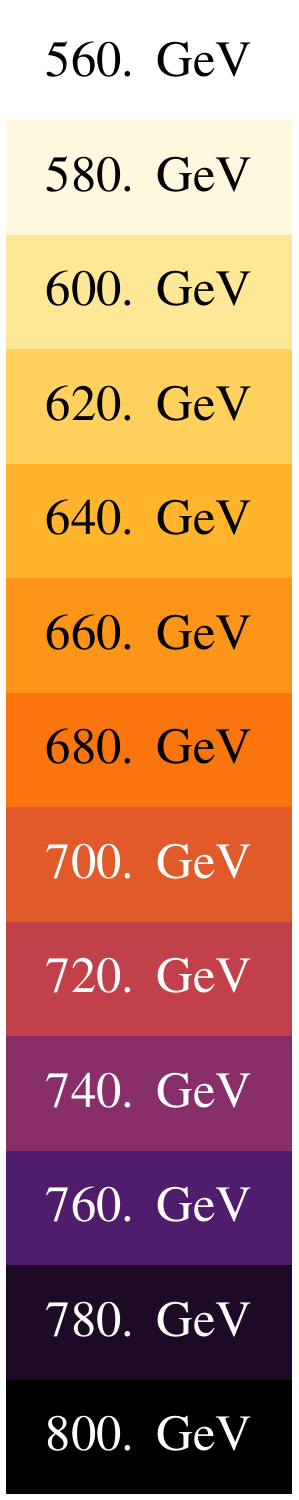}
  \caption{Ternary plots showing the direct constraints from the CMS search for heavy top quarks, for arbitrary combinations of $BR(T \rightarrow tH)$, $BR(T \rightarrow tZ)$, $BR(T \rightarrow bW)$, and $X= BR(T \rightarrow t\eta^{0}/b\eta^\pm) = 1 - BR(T \rightarrow tH) - BR(T \rightarrow tZ) - BR(T \rightarrow bW)$. Each branching ratio has a maximum value at the labeled corner, with a branching ratio of 0 on the opposing side. White markers indicate the progress of the simplified model branching ratio location for varying $s_L$, where the end points ($s_L=0.01$ and $s_L=0.59$) have been labeled. }
\label{fig:ternary}
\end{figure}

The direct search constraints are summarized in Figure \ref{fig:ternary} as ternary plots for five different values of $X=BR(T \rightarrow t\eta^{0}/b\eta^\pm)$, for all possible combinations of the other three possible decay modes ($bW$, $tZ$, $th^{0}$). The couplings of the heavy top partner to the Higgs and the electroweak gauge bosons as a function of the mixing between the SM top and the heavy top, $s_{L}$ is depicted by the white solid dots. Furthermore, the relationship between the $(bW,tZ,th^{0})$ decay modes does not change as the value of X increases. In addition, one can see from the figures, which represent slices of a tetrahedron, that as the value of $X$ increases, the decay of the heavy top is dominated by a single channel. This results in a scenario excluded at a fixed value of the heavy top mass for any combination of the original three decay modes, ($bW$, $tZ$, $th^{0}$). 

\section{Results}\label{sec:Results}
\begin{figure}[!h]
	\includegraphics[width=0.325\textwidth]{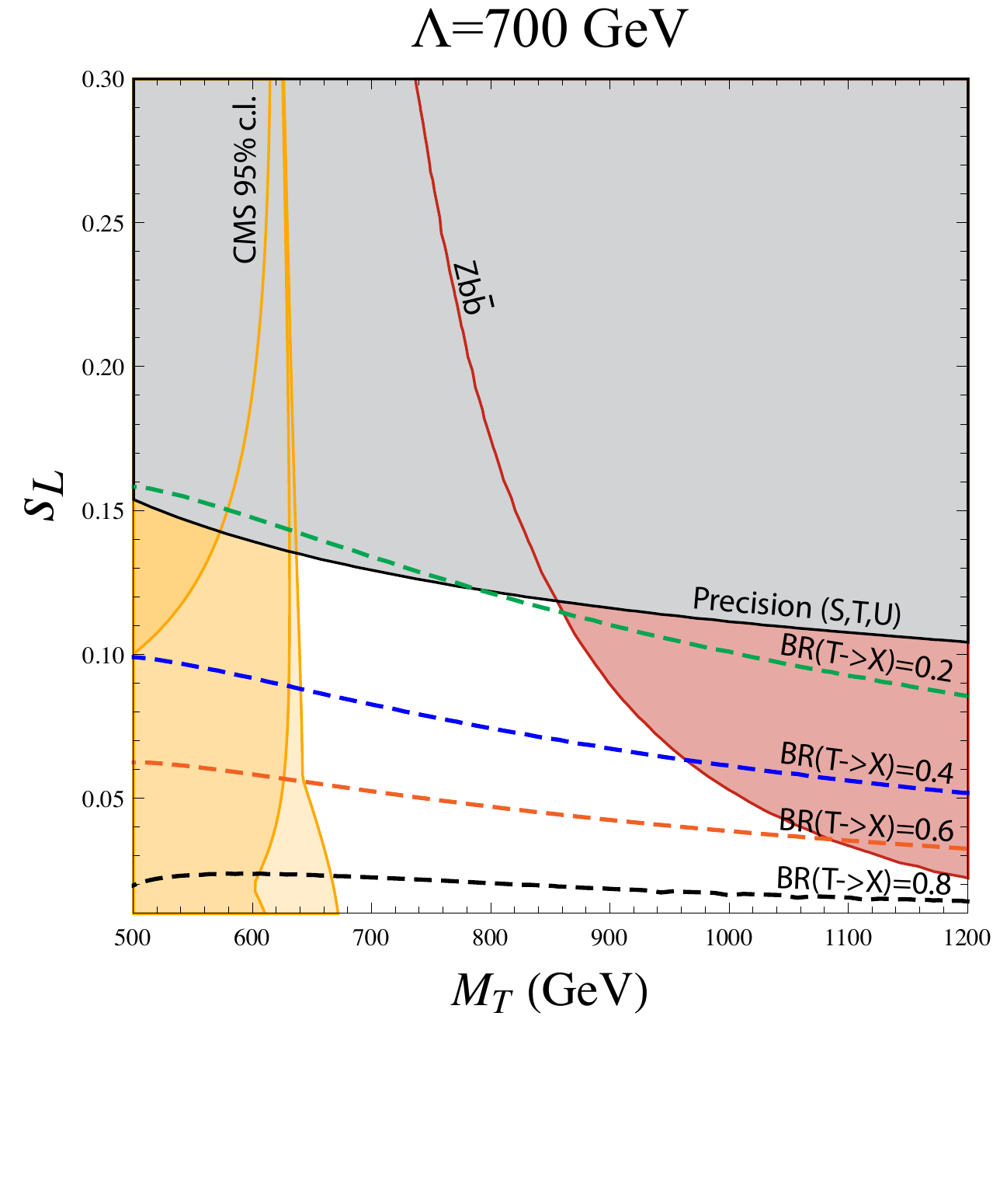}
	\includegraphics[width=0.325\textwidth]{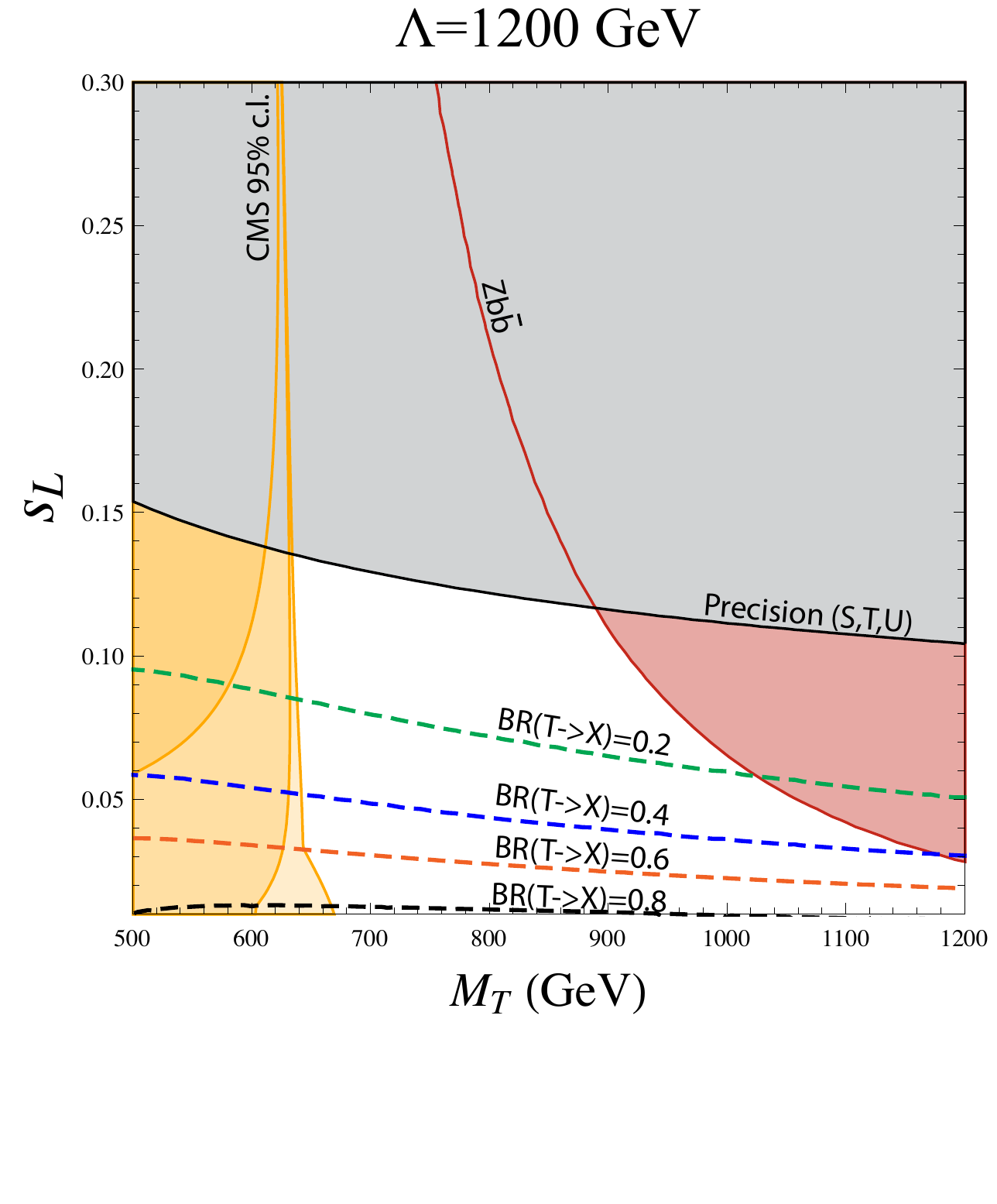}
	\includegraphics[width=0.325\textwidth]{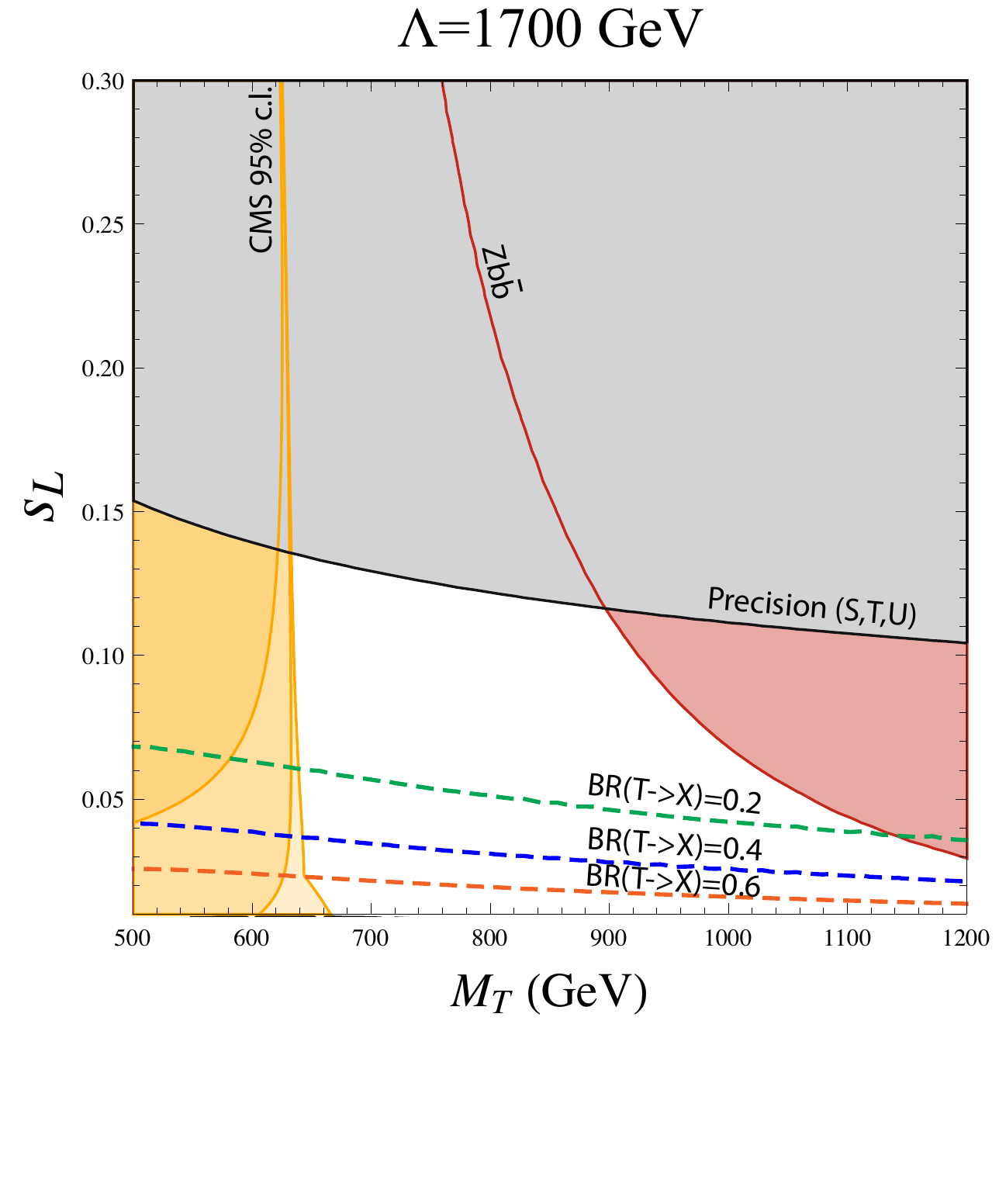}
	\vspace{-10mm}
\caption{\small The triplet + vector-like top quark model in the $s_{L}-m_{T}$ plane corresponding to three values of $\Lambda$ and using $y_{5}=1$. The grey and red regions are excluded by the $T$ observable and the $Zb\bar{b}$ vertex, respectively. The yellow region bounded by the outermost solid yellow line is excluded by CMS search for $T\bar{T}$ production discussed in the text. The dashed green, blue, orange, and black contours correspond to heavy top branching ratios of $0.2,0.4,0.6$ and $0.8$ respectively. (Labels included in figure for greyscale version.)} \label{fig:ToyResults}
\end{figure}


Using the couplings introduced in Equation~(\ref{eq:NewDecModes1}) we can calculate the branching ratios of a heavy vector-like top quark within the toy model introduced in Section~\ref{sec:tripletvector}. In order to extract the limits on the heavy top quark mass, we vary the left-handed mixing angle, $s_{L}$, and the mass of the heavy top, $m_{T}$. Furthermore, we analyze our model for different values of $\Lambda$ and fix $y_{5}=1$. The values of $\epsilon_{1}$ and $\epsilon_{2}$ are fixed to $2.5$. We assume that $\epsilon_{b}\approx y_{b}$ such that contributions to $g^{SM}_{R}$ are negligible. In this way, we fix all parameters to reasonable values. We impose the constraint on the top quark mass, the condition that leads to a solution to the Hierarchy Problem and choose to vary only $s_L$ and $m_T$. The results are shown in Figure~\ref{fig:ToyResults} for three values of $\Lambda$. Within the figure, the grey and red regions are excluded by the $T$ observable and the measurement of the $Zb\bar{b}$ vertex, respectively. The dashed green, blue, orange, and black contours correspond to heavy top branching ratios of $0.2,0.4,0.6$ and $0.8$ respectively. 

The two lightest yellow regions describes the exclusions from the CMS search for $T\bar{T}$ production when applied to model examined in this paper, where the lighter region corresponds to the model as described and the middle region corresponds to the situation where the scalar triplet masses are too heavy to allow the additional decay channels. The darkest yellow region describes the region excluded by the CMS search when we assume the $\eta$ decay products of the $T$ are not identified by the detector; this region is included purely for contrast. We see in all three cases of $\Lambda$ that the appearance of new decays modes of the heavy vector-like quark rules out a slightly larger area of parameter space, but approached a standard three-decay mode scenario as $\Lambda$ increases. Furthermore, for $\Lambda=700$ GeV; we obtain values of $y_{4}$ that are greater than one for vector-like masses, $m_{T}$, above $850$ GeV putting in question the validity of the effective model. Larger values of $\Lambda$ result in suppression of the $T\bar{t}\eta^0$ and $T\bar{b}\eta^-$ couplings, driving the branching ratio to the new states down. In addition, values of $\Lambda$ above $1$ TeV yield values of $y_{4}$ below unity in the region of $s_{L}$ consistent with experimental constraints and for vector-like masses below $1.2$ TeV.

Furthermore, in all three cases, a branching ratio to a new decay mode can be as large as $20\%$ when mixing between the SM top quark and the vector-like quark is small. This may serve as motivation for an in depth search that includes a decay mode corresponding to three top quarks at the LHC or incorporating $b-$tagging in future LHC searches, which would be a signature of a model that couples to fermions with Yukawa-like strengths.

\begin{figure}[!h]
\begin{center}
\includegraphics[width=0.5 \textwidth]{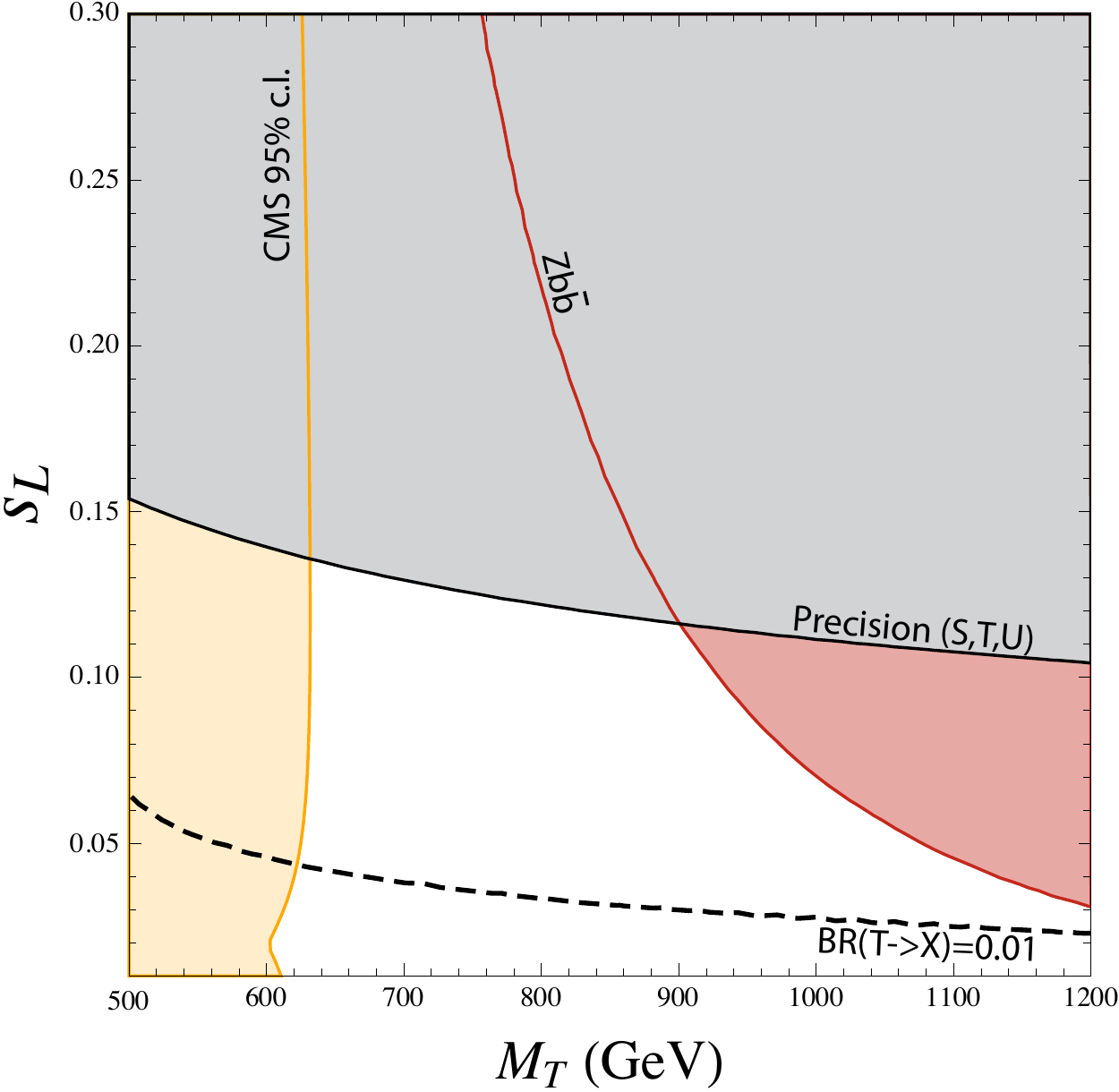}
\caption{\small The triplet + vector-like top quark model in the $s_{L}-m_{T}$ plane corresponding to $\Lambda=700$ GeV and $y_{5}=1$. The grey and red regions are excluded by the $T$ observable and the $Zb\bar{b}$ vertex. The yellow region bounded by the outermost solid yellow line is excluded by CMS search for $T\bar{T}$ production discussed in the text. The dashed line corresponds to a heavy top branching ratio of $1\%$.}
\label{fig:ToySym}
\end{center}
\end{figure}

The above results were generated by fixing the parameters $\epsilon_{1,2}$ to 2.5, enhancing the partial widths of the vector-like top quark to the real triple scalar for not too large values of $\Lambda$. However, it is interesting to analyze the case where the parameters $\epsilon_{1,2}$ are related to $y_{1,2}$ respectively. This relation is not unnatural since it may be the result of a more fundamental symmetry relating the couplings in the fermion Lagrangian, as will be seen in the next section. In our study, the values of $y_{1}$ and $y_{2}$ are found by fixing the values of the top quark mass to $173$ GeV as well as the heavy top mass using Equation~(\ref{eq:masses}). The results are shown in Figure~\ref{fig:ToySym}, where we have fixed  the value of $\Lambda$ to $700$ GeV, $y_{5}=1$ and used $\epsilon_{1,2}=y_{1,2}$. In the figure, the black dashed line corresponds to a branching ratio of the vector-like quark to the new scalar modes of $1\%$. Thus, the main decay modes of the vector-like top quark are the modes studied in minimal vector-like extensions of the SM, $BR(T\to th^{0},~tZ,~bW)$. This is clear since the existence of a new decay mode with a very small branching ratio is indistinguishable from the case of a decoupled scalar triplet or when the scalar triplet cannot be identified by the detector for $BR(T\to th^{0})+BR(T\to tZ)+BR(T\to bW)\approx1$. The smallness of the new branching ratio is mainly due to the fact that for small mixing angles, $s_{L}$, not ruled out by the $T$-parameter, $y_{2}$ tends to be small. 

\section{Conclusions}\label{sec:Conclusions}

Vector-like quark extensions of the SM have been extensively studied as a solution to the hierarchy problem. In particular, models where symmetries relating vector-like quarks to SM fermions are one attractive scenario. In this work, we have studied the phenomenology of a model where an extended scalar sector is coupled to the SM fermion sector and one single vector-like partner of the top quark. We have introduced new non-renormalizable interactions parametrized by the scale $\Lambda$ where new physics is expected to appear and at the same time used operators of the form $H^{\dagger}H\bar{Q}Q$ to address the electroweak hierarchy problem which many believe should be discoverable at the LHC. 

Within this framework, we have studied new decay modes of the heavy vector-like top quark which arise as a consequence of an extended fermion Lagrangian. These new modes are the neutral and charged components of a real scalar triplet in association with SM particles, mainly third generation up- and down-type quarks. We found that for couplings $\epsilon_{1}$ and $\epsilon_{2}$ that parametrize the new interactions between the scalar triplet and quarks and a new physics scale given by $\Lambda\sim1$ TeV, branching ratios to the scalar modes could be large and be consistent with electroweak precision measurements as well as the latest collider constraints on heavy vector-like pair production. However, we also found that equating $\epsilon_{1,2}$ to the couplings that parametrize the renormalizable interactions in the fermion Lagrangian, $y_{1,2}$ respectively, lead to large suppressions of the new decay modes when compared to those that appear in minimal vector-like extensions of the SM, $BR(T\to th^{0},~,tZ,~bW)$. In regards to the former case, our results serve as a motivation for an in depth search for final states corresponding to a large multiplicity of top quarks at the LHC. 

Within our model, the new scalar contributions to the $Zb\bar{b}$ vertex interfere with contributions from the heavy top. However, for a fixed mass of the real triplet, contributions to the $Zb\bar{b}$ vertex are strongest for large values of the heavy top mass since the loop functions depend quadratically on the mass of the heavy top. This has the effect of decreasing the excluded region of parameter space for large values of $s_{L}$ and small values of $m_{T}$. In addition, we found that the constraints from the $T$-parameter are unchanged from the scenario where only a $SU(2)_{W}$ singlet vector-like quark appears in the spectrum, since the contribution to the $T$-parameter from an inert real scalar triplet is negligible. These three constraint regions significantly limit the allowable region for this type of model for $s_L \gtrsim 0.03$. It is unlikely that enhancements in the $T$-parameter and $Zb\bar{b}$ constraints will occur in a the near future. However, we expect that the LHC13/14 program will be able to significantly increase the $T\bar{T}$ direct search limits, potentially even for masses upwards of 1 TeV.

Furthermore, we have also studied the possibility of embedding the toy model introduced in this study into a more fundamental framework, where the real scalar triplet and the heavy vector-like top quark are related to the SM particle content through global symmetries. In particular, we have studied the relationship between the parameters of our toy model to those within the Next to Littlest Higgs model. In this way, the toy model discussed in this work belongs to a low energy limit of the Next to Littlest Higgs model where the cancellation of the quadratic divergences in the Higgs mass arise naturally.

\section*{Acknowledgements}

The authors would like to thank Jernej Kamenik, David Morrissey, and John Ng for useful discussions and essential feedback regarding the progress of this work. This work is supported in parts by the National Science and Engineering Council of Canada.




\begin{thebibliography}{99}


\bibitem{Aad:2012tfa} 
  G.~Aad {\it et al.}  [ATLAS Collaboration],
  ``Observation of a new particle in the search for the Standard Model Higgs boson with the ATLAS detector at the LHC,''
  Phys.\ Lett.\ B {\bf 716}, 1 (2012)
  [arXiv:1207.7214 [hep-ex]].
  
\bibitem{Chatrchyan:2013lba} 
  S.~Chatrchyan {\it et al.}  [CMS Collaboration],
  ``Observation of a new boson with mass near 125 GeV in pp collisions at $\sqrt{s}$ = 7 and 8 TeV,''
  JHEP {\bf 1306}, 081 (2013)
  [arXiv:1303.4571 [hep-ex]].
  
  

  
\bibitem{ArkaniHamed:2002qy} 
  N.~Arkani-Hamed, A.~G.~Cohen, E.~Katz and A.~E.~Nelson,
  ``The Littlest Higgs,''
  JHEP {\bf 0207}, 034 (2002)
  [hep-ph/0206021].
  
\bibitem{Perelstein:2003wd} 
  M.~Perelstein, M.~E.~Peskin and A.~Pierce,
  ``Top quarks and electroweak symmetry breaking in little Higgs models,''
  Phys.\ Rev.\ D {\bf 69}, 075002 (2004)
  [hep-ph/0310039].
  

  
\bibitem{Contino:2006qr} 
  R.~Contino, L.~Da Rold and A.~Pomarol,
  ``Light custodians in natural composite Higgs models,''
  Phys.\ Rev.\ D {\bf 75}, 055014 (2007)
  [hep-ph/0612048].
  
\bibitem{Matsedonskyi:2012ym} 
  O.~Matsedonskyi, G.~Panico and A.~Wulzer,
  ``Light Top Partners for a Light Composite Higgs,''
  JHEP {\bf 1301}, 164 (2013)
  [arXiv:1204.6333 [hep-ph]].
  

\bibitem{Schmaltz:2010ac} 
  M.~Schmaltz, D.~Stolarski and J.~Thaler,
  ``The Bestest Little Higgs,''
  JHEP {\bf 1009}, 018 (2010)
  [arXiv:1006.1356 [hep-ph]].
  
\bibitem{Martin:2013fta} 
  T.~A.~W.~Martin and A.~de la Puente,
  ``Darkening the Little Higgs,''
  Phys.\ Lett.\ B {\bf 727}, 443 (2013)
  [arXiv:1304.7835 [hep-ph]].
  

\bibitem{Gherghetta:2000qt} 
  T.~Gherghetta and A.~Pomarol,
  ``Bulk fields and supersymmetry in a slice of AdS,''
  Nucl.\ Phys.\ B {\bf 586}, 141 (2000)
  [hep-ph/0003129].
  
\bibitem{Grossman:1999ra} 
  Y.~Grossman and M.~Neubert,
  ``Neutrino masses and mixings in nonfactorizable geometry,''
  Phys.\ Lett.\ B {\bf 474}, 361 (2000)
  [hep-ph/9912408].
  
\bibitem{Huber:2000ie} 
  S.~J.~Huber and Q.~Shafi,
  ``Fermion masses, mixings and proton decay in a Randall-Sundrum model,''
  Phys.\ Lett.\ B {\bf 498}, 256 (2001)
  [hep-ph/0010195].
  
\bibitem{Huber:2003tu} 
  S.~J.~Huber,
  ``Flavor violation and warped geometry,''
  Nucl.\ Phys.\ B {\bf 666}, 269 (2003)
  [hep-ph/0303183].
  
\bibitem{Agashe:2003zs} 
  K.~Agashe, A.~Delgado, M.~J.~May and R.~Sundrum,
  ``RS1, custodial isospin and precision tests,''
  JHEP {\bf 0308}, 050 (2003)
  [hep-ph/0308036].
  
\bibitem{Agashe:2004bm} 
  K.~Agashe and G.~Servant,
  ``Baryon number in warped GUTs: Model building and (dark matter related) phenomenology,''
  JCAP {\bf 0502}, 002 (2005)
  [hep-ph/0411254].
  
\bibitem{Agashe:2004cp} 
  K.~Agashe, G.~Perez and A.~Soni,
  ``Flavor structure of warped extra dimension models,''
  Phys.\ Rev.\ D {\bf 71}, 016002 (2005)
  [hep-ph/0408134].
  
\bibitem{Agashe:2006at} 
  K.~Agashe, R.~Contino, L.~Da Rold and A.~Pomarol,
  ``A Custodial symmetry for Zb anti-b,''
  Phys.\ Lett.\ B {\bf 641}, 62 (2006)
  [hep-ph/0605341].

  

\bibitem{Peskin:1991sw} 
  M.~E.~Peskin and T.~Takeuchi,
  ``Estimation of oblique electroweak corrections,''
  Phys.\ Rev.\ D {\bf 46}, 381 (1992).
  

\bibitem{Baak:2012kk} 
  M.~Baak, M.~Goebel, J.~Haller, A.~Hoecker, D.~Kennedy, R.~Kogler, K.~Moenig and M.~Schott {\it et al.},
  Eur.\ Phys.\ J.\ C {\bf 72}, 2205 (2012)
  [arXiv:1209.2716 [hep-ph]].
  
  
  
  

\bibitem{cmsHT}
  S.~Chatrchyan {\it et al.}  [CMS Collaboration],
  ``Inclusive search for a vector-like T quark with charge 2/3 in pp collisions at $\sqrt{s}$=8 TeV,''
  arXiv:1311.7667 [hep-ex].
  
\bibitem{atlasHT}
  The ATLAS collaboration,
  ``Search for pair production of heavy top-like quarks decaying to a high-$p_{\rm T}$ $W$ boson and a $b$ quark in the lepton plus jets final state in $pp$ collisions at $\sqrt{s}=8$ TeV with the ATLAS detector,''
  ATLAS-CONF-2013-060.
  
  





\bibitem{Gunion:1989ci} 
  J.~F.~Gunion, R.~Vega and J.~Wudka,
  ``Higgs triplets in the standard model,''
  Phys.\ Rev.\ D {\bf 42}, 1673 (1990).
  
\bibitem{Blank:1997qa} 
  T.~Blank and W.~Hollik,
  ``Precision observables in SU(2) x U(1) models with an additional Higgs triplet,''
  Nucl.\ Phys.\ B {\bf 514}, 113 (1998)
  [hep-ph/9703392].
  
\bibitem{Forshaw:2001xq} 
  J.~R.~Forshaw, D.~A.~Ross and B.~E.~White,
  ``Higgs mass bounds in a triplet model,''
  JHEP {\bf 0110}, 007 (2001)
  [hep-ph/0107232].
  
\bibitem{Forshaw:2003kh} 
  J.~R.~Forshaw, A.~Sabio Vera and B.~E.~White,
  ``Mass bounds in a model with a triplet Higgs,''
  JHEP {\bf 0306}, 059 (2003)
  [hep-ph/0302256].
  
\bibitem{Chen:2006pb} 
  M.~-C.~Chen, S.~Dawson and T.~Krupovnickas,
  ``Higgs triplets and limits from precision measurements,''
  Phys.\ Rev.\ D {\bf 74}, 035001 (2006)
  [hep-ph/0604102].
  
\bibitem{Chankowski:2006hs} 
  P.~H.~Chankowski, S.~Pokorski and J.~Wagner,
  ``(Non)decoupling of the Higgs triplet effects,''
  Eur.\ Phys.\ J.\ C {\bf 50}, 919 (2007)
  [hep-ph/0605302].
  
\bibitem{SekharChivukula:2007gi} 
  R.~S.~Chivukula, N.~D.~Christensen and E.~H.~Simmons,
  ``Low-energy effective theory, unitarity, and non-decoupling behavior in a model with heavy Higgs-triplet fields,''
  Phys.\ Rev.\ D {\bf 77}, 035001 (2008)
  [arXiv:0712.0546 [hep-ph]].
  

\bibitem{0811.3957}
  P.~Fileviez Perez, H.~H.~Patel, M.~.J.~Ramsey-Musolf and K.~Wang,
  ``Triplet Scalars and Dark Matter at the LHC,''
  Phys.\ Rev.\ D {\bf 79} (2009) 055024
  [arXiv:0811.3957 [hep-ph]].
  
  
  
  


\bibitem{1303.4490}
  L.~Wang and X.~-F.~Han,
  ``LHC diphoton Higgs signal in the Higgs triplet model with Y=0,''
  arXiv:1303.4490 [hep-ph].
  
\bibitem{Brdar:2013iea} 
  V.~Brdar, I.~Picek and B.~Radovcic,
  ``Radiative Neutrino Mass with Scotogenic Scalar Triplet,''
  Phys.\ Lett.\ B {\bf 728}, 198 (2014)
  [arXiv:1310.3183 [hep-ph]].
  
  
  
  
  
\bibitem{Cirelli:2005uq} 
  M.~Cirelli, N.~Fornengo and A.~Strumia,
  ``Minimal dark matter,''
  Nucl.\ Phys.\ B {\bf 753}, 178 (2006)
  [hep-ph/0512090].

  
\bibitem{9906332} 
  H.~E.~Logan,
  ``Radiative corrections to the $Zb\bar{b}$ vertex and constraints on extended Higgs sectors,''
  hep-ph/9906332.
  
 
  
  
  
  
  


\bibitem{AguilarSaavedra:2002kr} 
  J.~A.~Aguilar-Saavedra,
  Phys.\ Rev.\ D {\bf 67}, 035003 (2003)
  [Erratum-ibid.\ D {\bf 69}, 099901 (2004)]
  [hep-ph/0210112].
  
\bibitem{Botella:2012ju} 
  F.~J.~Botella, G.~C.~Branco and M.~Nebot,
  ``The Hunt for New Physics in the Flavour Sector with up vector-like quarks,''
  JHEP {\bf 1212}, 040 (2012)
  [arXiv:1207.4440 [hep-ph]].
  
\bibitem{Fajfer:2013wca} 
  S.~Fajfer, A.~Greljo, J.~F.~Kamenik and I.~Mustac,
  ``Light Higgs and Vector-like Quarks without Prejudice,''
  JHEP {\bf 1307}, 155 (2013)
  [arXiv:1304.4219 [hep-ph]].
  
\bibitem{Aguilar-Saavedra:2013wba} 
  J.~A.~Aguilar-Saavedra,
  ``Mixing with vector-like quarks: constraints and expectations,''
  EPJ Web Conf.\  {\bf 60}, 16012 (2013)
  [arXiv:1306.4432 [hep-ph]].
  
\bibitem{Botella:2013bsa} 
  F.~J.~Botella, M.~Nebot and G.~C.~Branco,
  ``Vector-like quarks and New Physics in the flavour sector,''
  J.\ Phys.\ Conf.\ Ser.\  {\bf 447}, 012061 (2013).
  
  
\bibitem{Tevatron} 
  T.~Aaltonen {\it et al.}  [CDF Collaboration],
  ``Search for a Heavy Top-Like Quark in $p\bar{p}$ Collisions at ${\surd}s = 1.96$~TeV,''
  Phys.\ Rev.\ Lett.\  {\bf 107}, 261801 (2011)
  [arXiv:1107.3875 [hep-ex]].
  

\bibitem{Berger:2012ec}
  J.~Berger, J.~Hubisz and M.~Perelstein,
  ``A Fermionic Top Partner: Naturalness and the LHC,''
  JHEP {\bf 1207} (2012) 016
  [arXiv:1205.0013 [hep-ph]].
  
\bibitem{Wang:2012gm} 
  L.~Wang and X.~F.~Han,
  ``The recent Higgs boson data and Higgs triplet model with vector-like quark,''
  Phys.\ Rev.\ D {\bf 86}, 095007 (2012)
  [arXiv:1206.1673 [hep-ph]].
  
\bibitem{Kearney:2013cca} 
  J.~Kearney, A.~Pierce and J.~Thaler,
  ``Exotic Top Partners and Little Higgs,''
  JHEP {\bf 1310}, 230 (2013)
  [arXiv:1306.4314 [hep-ph]].



\bibitem{Kearney:2013oia}
  J.~Kearney, A.~Pierce and J.~Thaler,
  ``Top Partner Probes of Extended Higgs Sectors,''
  JHEP {\bf 1308} (2013) 130
  [arXiv:1304.4233, arXiv:1304.4233 [hep-ph]].

\bibitem{Fukano:2013aea}
  H.~S.~Fukano, M.~Kurachi, S.~Matsuzaki and K.~Yamawaki,
  ``Higgs as a Top-Mode Pseudo,''
  arXiv:1311.6629 [hep-ph].




\bibitem{Gillioz:2013pba}
  M.~Gillioz, R.~Gröber, A.~Kapuvari and M.~Mühlleitner,
  ``Vector-like Bottom Quarks in Composite Higgs Models,''
  JHEP {\bf 1403} (2014) 037
  [arXiv:1311.4453 [hep-ph]].

\bibitem{Xiao:2014kba}
  M.~-L.~Xiao and J.~-H.~Yu,
  ``Stabilizing Electroweak Vacuum in a Vector-like Fermion Model,''
  arXiv:1404.0681 [hep-ph].

\bibitem{Karabacak:2014nca}
  D.~Karabacak, S.~Nandi and S.~K.~Rai,
  ``New signals for singlet Higgs and vector-like quarks at the LHC,''
  arXiv:1405.0476 [hep-ph].
  
\bibitem{Bahrami:2014ska}
  S.~Bahrami and M.~Frank,
  ``Vector Quarks in the Higgs Triplet Model,''
  arXiv:1405.4245 [hep-ph].


  

  
  
  
\bibitem{Blankenburg:2012ex} 
  G.~Blankenburg, J.~Ellis and G.~Isidori,
  ``Flavour-Changing Decays of a 125 GeV Higgs-like Particle,''
  Phys.\ Lett.\ B {\bf 712}, 386 (2012)
  [arXiv:1202.5704 [hep-ph]].
  
\bibitem{Harnik:2012pb} 
  R.~Harnik, J.~Kopp and J.~Zupan,
  ``Flavor Violating Higgs Decays,''
  JHEP {\bf 1303}, 026 (2013)
  [arXiv:1209.1397 [hep-ph]].
  
  
  
  

\bibitem{GenericSVQ} 
  J.~A.~Aguilar-Saavedra, R.~Benbrik, S.~Heinemeyer and M.~Pérez-Victoria,
  ``Handbook of vectorlike quarks: Mixing and single production,''
  Phys.\ Rev.\ D {\bf 88}, no. 9, 094010 (2013)
  [arXiv:1306.0572 [hep-ph]].
  
  
  
  
\bibitem{CapdequiPeyranere:1990qk} 
  M.~Capdequi Peyranere, H.~E.~Haber and P.~Irulegui,
  ``$H^+- \to W^+ \gamma$ and $H+- \to W^+- Z$ in two Higgs doublet models. 1. The Large fermion mass limit,''
  Phys.\ Rev.\ D {\bf 44}, 191 (1991).
  
  
  
  
  
  

  
\bibitem{ALEPH:2005ab} 
  S.~Schael {\it et al.}  [ALEPH and DELPHI and L3 and OPAL and SLD and LEP Electroweak Working Group and SLD Electroweak Group and SLD Heavy Flavour Group Collaborations],
  ``Precision electroweak measurements on the $Z$ resonance,''
  Phys.\ Rept.\  {\bf 427}, 257 (2006)
  [hep-ex/0509008].
  
\bibitem{Bsmumu} 
  D.~Guadagnoli and G.~Isidori,
  ``$B(B_s \to \mu^{+} \mu^{-})$ as an electroweak precision test,''
  Phys.\ Lett.\ B {\bf 724}, 63 (2013)
  [arXiv:1302.3909 [hep-ph]].
  
  
  
  
  
\bibitem{Barducci:2014ila} 
  D.~Barducci, A.~Belyaev, M.~Buchkremer, G.~Cacciapaglia, A.~Deandrea, S.~De Curtis, J.~Marrouche and S.~Moretti {\it et al.},
  ``Model Independent Framework for Analysis of Scenarios with Multiple Heavy Extra Quarks,''
  arXiv:1405.0737 [hep-ph].
  
\bibitem{Berger:2009qy} 
  E.~L.~Berger and Q.~-H.~Cao,
  ``Next-to-Leading Order Cross Sections for New Heavy Fermion Production at Hadron Colliders,''
  Phys.\ Rev.\ D {\bf 81}, 035006 (2010)
  [arXiv:0909.3555 [hep-ph]].
  
  
\bibitem{HATHOR} 
  M.~Aliev, H.~Lacker, U.~Langenfeld, S.~Moch, P.~Uwer and M.~Wiedermann,
  ``HATHOR: HAdronic Top and Heavy quarks crOss section calculatoR,''
  Comput.\ Phys.\ Commun.\  {\bf 182}, 1034 (2011)
  [arXiv:1007.1327 [hep-ph]].
  
\bibitem{checkmate} 
  M.~Drees, H.~Dreiner, D.~Schmeier, J.~Tattersall and J.~S.~Kim,
  arXiv:1312.2591 [hep-ph].







%
%
%
%




\bibitem{Deshpande:1977rw} 
  N.~G.~Deshpande and E.~Ma,
  ``Pattern of Symmetry Breaking with Two Higgs Doublets,''
  Phys.\ Rev.\ D {\bf 18}, 2574 (1978).
  
\bibitem{LopezHonorez:2006gr} 
  L.~Lopez Honorez, E.~Nezri, J.~F.~Oliver, M.~H.~G.~Tytgat and ,
  ``The Inert Doublet Model: An Archetype for Dark Matter,''
  JCAP {\bf 0702}, 028 (2007)
  [hep-ph/0612275].

\bibitem{Dolle:2009fn} 
  E.~M.~Dolle, S.~Su and ,
  ``The Inert Dark Matter,''
  Phys.\ Rev.\ D {\bf 80}, 055012 (2009)
  [arXiv:0906.1609 [hep-ph]].
  
\bibitem{Pappadopulo:2010jx} 
  D.~Pappadopulo and A.~Vichi,
  ``T-parity, its problems and their solution,''
  JHEP {\bf 1103}, 072 (2011)
  [arXiv:1007.4807 [hep-ph]].

\bibitem{Brown:2010ke} 
An inert doublet potential is also present in the $SU(6)/Sp(6)$ model described in T.~Brown, C.~Frugiuele and T.~Gregoire,  
  JHEP {\bf 1106}, 108 (2011)  [arXiv:1012.2060 [hep-ph]]. Our assertion is that the inert doublet potential can be generated within many little Higgs models, while also relieving precision and fine-tuning constraints, through the modification we propose.
  
  
  



\end{thebibliography}
\end{document}